\documentclass[final,12pt]{article}

\RequirePackage{amsthm,amsmath,amsfonts,amssymb}
\RequirePackage[authoryear]{natbib}
\RequirePackage[colorlinks,citecolor=blue,urlcolor=blue]{hyperref}
\RequirePackage{color,graphicx}

\usepackage{subfigure}
\usepackage{booktabs}
\usepackage{times}
\usepackage{float}
\usepackage[margin=1in]{geometry}
\usepackage{mleftright}
\usepackage{mathtools}
\usepackage{ gensymb }
\usepackage{ stmaryrd }
\usepackage{authblk}
\usepackage{csquotes}
\usepackage{lipsum}

\MakeOuterQuote{"}

\newcommand{\red}[1]{\textcolor{red}{#1}}

\graphicspath{ {./Figures/} }

\providecommand{\U}[1]{\protect \rule{.1in}{.1in}}
\newtheorem{theorem}{Theorem}
\AfterEndEnvironment{theorem}{\noindent\ignorespaces}

\newtheorem{definition}{Definition}
\newtheorem{lemma}{Lemma}
\AfterEndEnvironment{lemma}{\noindent\ignorespaces}

\title{Optimal Conformal Prediction for Small Areas}

\author{Elizabeth Bersson}
\author{Peter D. Hoff}
\affil{Department of Statistical Science, Duke University}

\date{\today}

\newcommand\reals{\mathbb{R}}

\usepackage{bbm}

\newcommand\variance{\sigma^2}

\newcommand\bs{\Sigma}

\newcommand\onen{\mathbf{1}_n}
\newcommand\onent{\mathbf{1}_n^T}

\newcommand\one{\mathbf{1}}

\newcommand\thetabf{{\boldsymbol{\theta}}}
\newcommand\Ybf{\boldsymbol{Y}}
\newcommand\wbf{\boldsymbol{w}}
\newcommand\ybf{\boldsymbol{y}}

\newcommand\distiid{\stackrel{iid}{\sim}}

\begin{document}

\maketitle

\begin{abstract}
Existing inferential methods for small area data involve a trade-off between maintaining area-level frequentist coverage rates and improving inferential precision via the incorporation of indirect information. In this article, we propose a method to obtain an area-level prediction region for a future observation which mitigates this trade-off. The proposed method takes a conformal prediction approach in which the conformity measure is the posterior predictive density of a working model that incorporates indirect information. The resulting prediction region has guaranteed frequentist coverage regardless of the working model, and, if the working model assumptions are accurate, the region has minimum expected volume compared to other regions with the same coverage rate. When constructed under a normal working model, we prove such a prediction region is an interval and construct an efficient algorithm to obtain the exact interval. We illustrate the performance of our method through simulation studies and an application to EPA radon survey data.

\smallskip \noindent \textit{Keywords}:  
Exchangeability; Hierarchical model; Kriging; Nonparametric; Prediction region; Frequentist coverage.

\end{abstract}

\section{Introduction}\label{intro}

Precise and accurate inference on a sample obtained from non-overlapping subpopulations, referred to as areas or domains, is an important goal in a wide range of fields including 
economics \citep{Berg2014}, ecology \citep{Sinha2009}, and others  \citep{Chattopadhyay1999}  where localized inference for various socio-demographic groups or refined geographic regions
is of interest. 
In such applications, it is common to have small area-specific sample sizes, which presents
challenges in making precise area-specific inferences via direct methods that only make use of within-area samples. 
Direct methods can be unbiased and achieve target frequentist coverage rates for all areas, but 
don't take advantage of auxiliary information,
and so may be inefficient. 
As a result, researchers often turn to indirect or model-based methods that allow information to be shared across areas. 
Borrowing information across areas may decrease 
variability of point estimates and volume of confidence and prediction regions, but doing so 
can introduce bias and thus alter area-level frequentist coverage rates from their nominal level \citep{Carlin1990}.
For more on small area inference, see \cite{Rao2015}, 
or, for information on multilevel modeling more broadly, see
\cite{Gelman2006a}. 
Recently, some parametric `frequentist and Bayes' (FAB) methods have been developed for confidence intervals that
maintain area-level frequentist error rate control and allow for information sharing \citep{Yu2018,Burris2020}. 

In this article,
we develop a non-parametric FAB prediction method that 
has area-level frequentist coverage rate control while
incorporating indirect information to improve prediction region precision.
Specifically, we focus on
the task 
of predicting a future response in each small area.
Although it is not the focus of much of the mixed effects model literature, response prediction is particularly useful for small areas.
It is more general than quantifying area-level effects and allows for a more natural comparison across different types or levels of groups.
Additionally, it may allow for more straightforward interpretation of differences across areas than estimates of effects, particularly in non-normal populations.
More broadly, the underlying goal of most statistical inference can be 
framed as being motivated by
prediction \citep{Shmueli2010}, so it is natural to directly target this goal.

\begin{figure}[h]
\centering
\includegraphics[width=.5\textwidth,keepaspectratio]{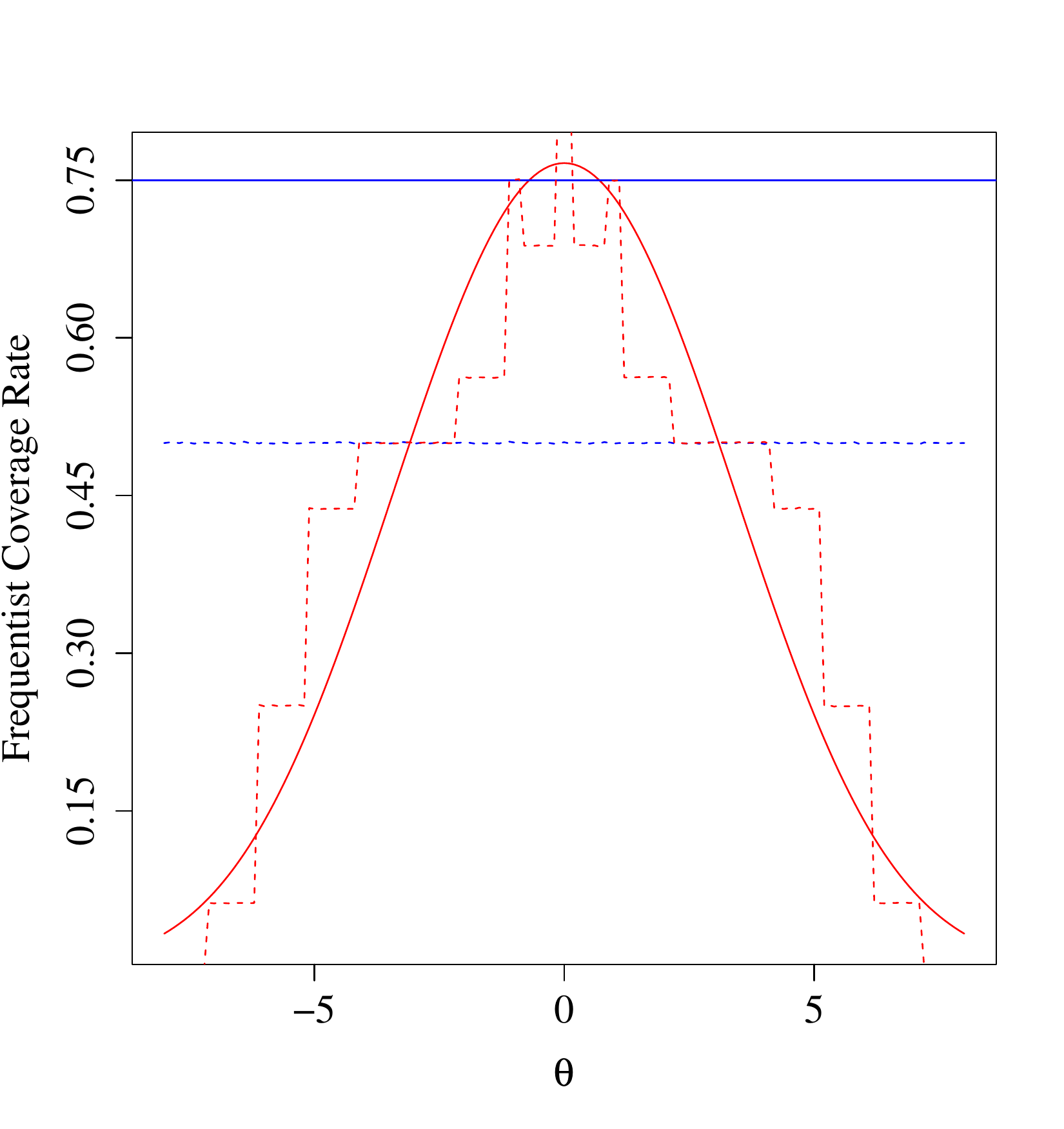}
\caption{
Frequentist coverage rate of classical pivot prediction (blue) and Bayesian prediction (red) for sample size $n=3$, known $\variance = 1$, and Bayesian prior parameters $\mu = 0,\tau^2=1/2$, for a target coverage rate of $0.75$.
Results when parametric assumptions are accurate
(solid lines) and 
inaccurate (dashed lines).
}\label{all_coverage}
\end{figure}

To illustrate the limitations of commonly used parametric prediction methods, 
consider a study that includes an exchangeable sample 
$\boldsymbol Y_j = \{ Y_{1,j} ,\ldots, Y_{n_j,j}\}$
obtained from $J$ populations
such that the samples are independently distributed across the populations.
We wish to obtain a prediction region $A_j$ for a future response from population, or area, $j$, $Y_{n_j+1,j}$, that is accurate, in the sense that it maintains $(1-\alpha)100\%$ frequentist coverage,
\begin{equation}
P_\theta\left(Y_{n_j+1,j}\in A_j\right) = 1- \alpha,
\end{equation}
and precise, in that the expected volume is comparatively small.
For data that appear approximately normally distributed, a commonly used direct method is the classical normal or $t$ pivot prediction interval.
To see what can go wrong, consider the simple case where the population variance is known. In this case, the pivot interval is of the form 
\[
\bar{y}_j \pm z_{1-\alpha/2}\left({\sigma}_j^2(1+1/n_j)\right)^{1/2},
\]
where $\bar{y}_j$ is the sample mean of area $j$, ${\sigma}_j^2$ is the population variance, and $z_{q}$ is the $q$th quantile of the standard normal distribution. 
If the parametric distributional assumptions hold true, this interval will have the desired frequentist coverage rate. 
As this interval may be prohibitively wide due to a small sample size, 
researchers often turn to a
Bayesian interval,
guaranteed to be narrower than the pivot interval: 
\[
\tilde{\theta}_j \pm z_{1-\alpha/2}\left(\left(1/\tau^2 + n/{\sigma}_j^2\right)^{-1} + {\sigma}_j^2\right)^{1/2},
\]
where $\tilde{\theta}_j := \left(\mu/\tau^2+\bar{y}_j n/{\sigma}_j^2\right)/\left(1/\tau^2 + n_j/{\sigma}_j^2\right)$, for prior parameters $\mu$ and $\tau^2$ that may be estimated in an empirical Bayesian manner.
This interval achieves the specified coverage rate on average across groups, but the
frequentist coverage rate 
declines from greater than $1-\alpha$ when $\theta_j=\mu$ to zero as
$|\theta_j-\mu|$ increases (see the solid lines in Figure \ref{all_coverage}).
Furthermore, if the parametric assumptions underlying either of these methods are not accurate, 
the methods' respective guarantees of coverage fail even on average across groups.
Consider, for example, a sample from a single population arising from 
a distribution characterized by mean $\theta$ and variance $1$ with a point mass equidistant on either side of $\theta$:
\[
f_{y}(y) = \lambda \delta_{\theta-1}(y) +(1-\lambda)\delta_{\theta+1}(y),
\]
for $\lambda\sim Bernoulli(1/2)$ where $\delta_c$ is the Dirac delta function at point $c$. 
For such a population, depending on the specified $\alpha$,
the frequentist coverage rate of the standard pivot method
can be much lower than the desired rate,
and 
the frequentist coverage rate of the Bayesian prediction method 
can be greater or lower than expected depending
on $|\theta_j-\mu|$
(see dashed lines in Figure \ref{all_coverage}).
In summary, these popular parametric prediction methods incur a lower than desired frequentist coverage rate in the case of incorrect prior values or inaccurate parametric assumptions.
This behavior is particularly problematic when area-specific inference is specifically of interest as in small area applications,
where small sample sizes make forming accurate distributional assumptions difficult.

There is an extensive body of literature on estimating (or, often, "predicting") random effects for small areas (see, for a review, \cite{Skrondal2009} or \cite{Pfeffermann2013}), but less work has been done on response prediction. 
\cite{Afshartous2005} offer a review of parametric point prediction methods. 
The accuracy and theoretical guarantees of these methods rely on modeling assumptions.
Other prediction methods such as those presented in
\cite{Vidoni2006}
,
as well as empirical or fully Bayesian prediction methods
produce precise prediction intervals \citep{Gelman2006}, but do not maintain the desired coverage level at each area.

Conformal prediction,  introduced in \cite{Gammerman1998} and further developed in \cite{Vovk2005},
 is a non-parametric method 
which relys solely on the assumption of exchangeability to produce 
prediction regions guaranteed to have the desired coverage.
Candidate predicted values are included in a conformal prediction region 
if they are similar to the observed data,  as judged by a chosen conformity measure. A brief review of the conformal prediction method is included in Section \ref{conformal}. 
Much of the work on the conformal approach up to now has been
focused on 
methods for a single population. 
These methods 
\citep{Lei2013,Papadopoulos2011,Vovk2019}
may be used to construct "direct" conformal prediction regions for each area separately, but doing so could be inefficient, as information is not shared across groups. 


In this article, we propose a FAB conformal prediction method that utilizes indirect information. 
Our method results in prediction regions 
with nonparametric frequentist coverage guarantees that 
may have minimum expected volume
when 
precise indirect information is available.
Specifically, 
we build on the generic result shown in \cite{Hoff2021a} that 
conformal prediction regions obtained under the posterior predictive distribution as the conformity measure 
are optimally precise.
We show how this result can be used to obtain narrower prediction regions than standard methods by incorporating indirect information.
When the proposed conformity measure is constructed under a normal working model, we prove the resulting
prediction region is an interval that contains a standard Bayesian point prediction estimator. 
This implies a coherent method of classically Bayesian point prediction while providing uncertainty quantification which maintains frequentist coverage.
While many conformal prediction methods 
for complex conformity measures
rely on algorithms which 
result in approximate prediction regions, 
we develop a computationally efficient procedure that makes full use of the data to obtain the exact FAB conformal prediction region.

This article proceeds as follows. 
In Section \ref{mainconf}, we briefly review the generic conformal prediction method and
detail the motivation, properties, and computation of the Bayes-optimal, or FAB,  conformity measure for a single population when indirect information is available. 
Numerical results on the FAB prediction interval's precision are 
shared
in Section \ref{singlepopsim}.
In Sections \ref{fabsapmain} and \ref{radonexample}, we 
extend the FAB conformal algorithm to a small area or multiple population regime and illustrate the method's performance through an application to an EPA radon dataset.
We conclude with a discussion in Section \ref{concl}.
All proofs are contained in Appendix \ref{proofsall}.

\section{Bayes Optimal Conformal Prediction}\label{mainconf}

\subsection{Review of Conformal Prediction}\label{conformal}

Conformal prediction is a method of obtaining a prediction region for a new observation 
$Y_{n+1}$ based on an 
exchangeable sample $\Ybf=\{Y_1,...,Y_n\}$ from a real-valued population. 
Having observed $\Ybf = \boldsymbol{y}$, a candidate value $y_{n+1}$ of $Y_{n+1}$ 
is included in the conformal prediction region if 
it sufficiently ``conforms'' to the sample, as measured by a 
\emph{conformity measure} $C: \mathbb R^{n}\times \mathbb R \rightarrow \mathbb R$. 
The reader is referred to \cite{Vovk2005} \S 2.1-2.2 for more details. 
The conformal prediction region can be constructed to have the desired frequentist coverage 
rate by including only those $y_{n+1}$-values with corresponding conformity score $c_{n+1}$ greater than or equal to that of some fraction of the conformity scores of the observed  elements of the sample, $\{c_1,...,c_n\}$. 
Specifically, a $100 (1-\alpha)\%$ prediction region for $Y_{n+1}$ can 
be constructed as follows: 
To determine if a candidate value $y_{n+1}$ is 
included in the prediction region, 
\begin{enumerate}
\item  compute $c_i(y_{n+1}) := C\left(\{ y_1,..,y_n,y_{n+1}\}\backslash\{ y_i\},y_i\right)$  for $i=1,\ldots,n+1$; 
\item set $p_y := \frac{\text{$\# \big\{i=1,...,n+1:c_i(y_{n+1}) \leq c_{n+1}(y_{n+1}) \big\}$}}{n+1}$. 
\end{enumerate}
The value $y_{n+1}$ is included in the region if $p_y>\alpha$. Note that each conformity score $c_i(y_{n+1})$ is a function of the candidate $y_{n+1}$.
More compactly, the conformal prediction region may be expressed as
\begin{equation}\label{boilsdown}
A^c\left(\Ybf\right)= \left\{y_{n+1}\in\mathcal{Y}:\frac{\#\left\{i=1,...,n+1:c_i(y_{n+1}) \leq c_{n+1}(y_{n+1}) \right\}}{n+1}>\frac{k}{n+1}\right\},
\end{equation}
where $k = \lfloor\alpha(n+1)\rfloor$.
The resulting prediction region, $A^C(\Ybf)$, has greater than or equal to the desired coverage, referred to as conservative coverage,
and has exact coverage if $\epsilon = l/(n+1)$ for some integer $l\in\{0,1,2,...,n+1\}$.
The frequentist coverage guarantee follows from the exchangeability assumption as all permutations of the collection of random variables $\left\{ Y_1,...,Y_{n+1}\right\}$ are equiprobable, and thus, all permutations of conformity scores $\left\{ c_1,...,c_{n+1}\right\}$ are equiprobable 
(\cite{Balasubramanian2014} \S 1.3).

In summary, the conformal prediction method quantifies how well a candidate prediction conforms with previously observed data.  
If the candidate is more similar than some specified number of observations in the sample, it is included in the prediction interval.
Formally, the candidate is accepted if it's corresponding conformity score is greater than $k$ of the entire "bag" of conformity scores. 
A key advantage of the conformal algorithm is that the frequentist coverage guarantee of the conformal method holds regardless of both the true distribution of the random variables and the choice of conformity measure. 
(Thus, a thoughtfully chosen conformity measure
will yield a precise prediction region which maintains the desired coverage rate.)

\subsection{Bayes-Optimal Conformal Prediction via a Normal Working Model}\label{fabconfmain}

Two main criteria of the usefulness of a prediction region are validity and precision.
As frequentist validity is guaranteed by the conformal method, we focus on constructing an optimally precise prediction region through a conformity measure that takes advantage of auxiliary or prior information.
This information will enter the conformity measure through a \textit{working model}.
In practice, we often wish to make a prediction based on a random sample arising from some unknown distribution. 
There may be reason to believe the population arises from a working model, that is, some specific distribution that possibly incorporates prior information on model parameters.

In this section, we derive a  Bayes-optimal conformity measure $C_B$ for a single population using a normal working model:
\begin{align}\label{workingmodel}
 Y_1,...,Y_n\sim{}& Normal(\theta,\variance)\\
\theta\sim {}& Normal(\mu,\tau^2\variance)\nonumber\\
1/\variance\sim{}&  Gamma(a/2,b/2).\nonumber
\end{align}
The conformity measure is Bayes-optimal in the sense that it results in a prediction region 
with minimum expected volume for a specified coverage rate
if the working model is true.
Regardless of the accuracy of the working model, the resulting prediction region will maintain the desired coverage rate.
\cite{Hoff2021a} provides a generic result that the
Bayes-optimal conformity measure is
the posterior predictive density of the model:
\begin{equation}\label{fab}
C_{B}\left({\ybf},y_{n+1}\right) = p\left(y_{n+1}|\ybf\right) = \int_\Theta p(y_{n+1}|\thetabf)p(\thetabf | \ybf)d\thetabf,
\end{equation}
where $p(y_{n+1}|\thetabf)$ is the probability density of candidate $y_{n+1}$ given model parameters $\thetabf$, and $p(\thetabf|\ybf)$ is the posterior density of $\thetabf$ conditional on data $\ybf$.
We expand on this result to develop a method which makes use of indirect information 
to obtain narrower prediction regions than possible under standard methods. 
We derive specific results on
FAB conformal prediction under the normal working model.
Specifically, we prove properties of the resulting FAB conformal prediction region and develop an efficient computational algorithm to obtain the exact region.

We proceed with the derivation of the Bayes-optimal conformity measure under a normal working model.
A standard calculation shows that the posterior predictive density of the normal working model (Equations \ref{workingmodel}) is a non-standard, non-central $t$ density:
\begin{equation}\label{postpred}
p(y_{n+1}|\ybf) =
\frac{\Gamma \left(\frac{2a_\sigma+1}{2}\right)}{\sqrt{2a_\sigma\pi}\Gamma\left(\frac{2a_\sigma}{2}\right)} 
\left(\frac{1}{\sqrt{\variance_t }}
\left( 1  +\frac{1}{2a_\sigma}\frac{(y_{i}-\mu_\theta)^2}{\variance_t }\right)^{-(2a_\sigma+1)/2}\right) ,
\end{equation}
where 
\begin{align*}
{}& \tau^2_\theta = (1/\tau^2+n)^{-1}
{}&\mu_\theta = (\mu/\tau^2+\onent \ybf)\tau^2_\theta\\
{}&a_\sigma = a+n
{}&b_\sigma = b +\ybf^T\ybf+\mu^2/\tau^2-\left(\tau^2_\theta\right)^{-1}\mu_\theta^2
\end{align*}
and $\variance_t = \frac{b_\sigma}{a_\sigma}(1+\tau^2_\theta)$.
For now, we assume all hyperparameters of the working model,
$\{\mu,\tau^2,a,b\}$, are known.  In practice,  they may be estimated or otherwise obtained from auxiliary information. This will be more thoroughly discussed in the Section \ref{estprocedure}.
Given working model hyperparameters,
a FAB conformal prediction region, denoted $A^{fab}(\Ybf)$, can be constructed for a presumably normal population 
by taking Equation \ref{postpred} as the conformity measure.
Regardless of the accuracy of the working model, the resulting prediction region will maintain the specified coverage rate, and if the working model assumptions are accurate, the prediction region will 
have minimum expected volume. 
In what follows, we will derive properties and an efficient computational algorithm for the FAB conformal prediction region.

Given a conformity measure, it is straightforward to obtain the prediction region 
by implementing the conformal algorithm for each potential candidate prediction value in the sample space. 
In practice, this may be infeasible due to an infinite number of candidates, or otherwise computationally expensive to do at a meaningful granule. 
Alternatively, we may make use of the form of the chosen conformity measure to circumvent evaluating the typical conformal algorithm at each candidate value and obtain a more computationally tractable algorithm.
As we will show, computations may be facilitated by
considering alternative representations of a conformity measure, or ECMs:
\begin{definition}[equivalent conformity measure]
Two conformity measures are called equivalent conformity measures (ECM) if the resulting conformal prediction regions are equivalent.
\end{definition}
The idea of an ECM and 
how it may be used to simplify computation of the prediction region
have been discussed before
in the conformal literature.
For example, standard computation of the prediction region resulting from the popular distance to the average (DTA) non-conformity measure, $
C_{avg}(\ybf,y_{n+1}) = \left|y_{n+1}-\overline{\ybf}\right|,$
seems to require
computing the mean of $n+1$ sets
during the execution of the conformal algorithm.
As discussed in \cite{Shafer2008}, 
this can be avoided by using 
an ECM,
$C_{avg}\left(\{ \ybf,y_{n+1}\},y_{n+1}\right)$.
Considering this representation in place of the classically defined measure 
allows each conformity score to be defined in terms of the sample mean, an element of the sample, and the unknown candidate $y_{n+1}$.
This in turn simplifies the
implementation of the conformal algorithm.
Under more dynamic conformity measures such as the Bayes-optimal measure, the
computational gain obtained from constructing an algorithm under an ECM may be substantial.
It turns out that under the normal working model, 
the Bayes-optimal conformity measure has the same ECM property as the DTA measure:
\begin{theorem}\label{conformitytheorem}
Under Model \ref{workingmodel}, $C_{B}\left(\ybf,y_{n+1} \right)$ and $C_{B}\left(\{\ybf,y_{n+1}\},y_{n+1} \right)$ are ECM.
\end{theorem} 
Consideration of the ECM $C_{B}\left(\{\ybf,y_{n+1}\},y_{n+1} \right)$ greatly simplifies the computation of the FAB prediction region.
In the evaluation of the conformal algorithm under this measure, 
the conformity scores corresponding to each element of the sample and the candidate are each a $t$ density with the same parameters.
As such, this form of the conformity measure is more convenient to work with.

\subsection{Prediction Region Properties and Computation}\label{fabcomputation}

In order to obtain the conformal prediction region, in principle, 
the conformal algorithm must be evaluated for each candidate in the sample space $\mathcal{Y}$.
As such, unless $\mathcal{Y}$ is a finite set, the conformal prediction method could, in general, be prohibitively computationally expensive. 
However, by making use of properties of the form of $C_B$ under the normal working model (Equations \ref{workingmodel}), 
we show that
the exact conformal prediction region can be obtained by a procedure that involves
evaluating a simple function of the sample. 
Additionally, we prove that the FAB conformal prediction region under the normal working model will be an interval that contains the posterior mean estimator of the population mean, $\tilde{\theta} :=(\mu/\tau^2 + \sum_{k=1}^n y_k)(1/\tau^2+n)^{-1}$.

The FAB conformal prediction region
can be obtained
via a two step process. 
First, for each $i=1,...,n+1$, find the sub-region of acceptance,
\begin{equation}
S_i:=\left\{y_{n+1}\in\mathbb{R}:c_{B,i}(y_{n+1})\leq c_{B,n+1}(y_{n+1})\right\}.
\end{equation}
Then, information in the set $\{S_1,...,S_{n+1}\}$ can be summed over the domain $\reals$ to 
obtain the number of $i=1,...,n+1$ such that $c_i\leq c_{n+1}$ at each point in the domain. 
As made clear by the representation of a generic conformal prediction region given in Equation \ref{boilsdown}, this information fully classifies the conformal prediction region for a given error rate $\alpha$.

\begin{figure}[ht]
\centering
\includegraphics[width=.323\textwidth,keepaspectratio]{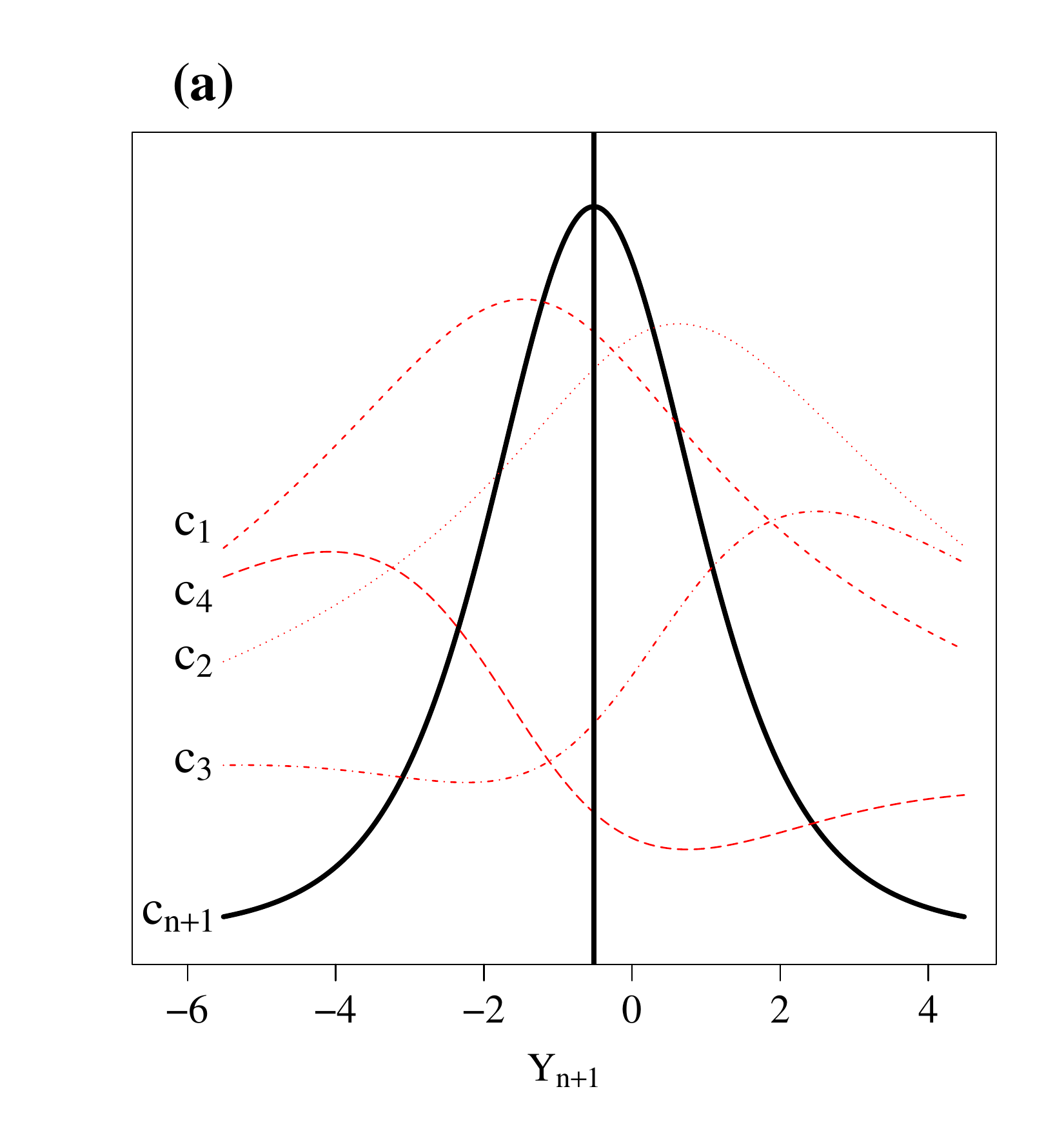} 
\includegraphics[width=.323\textwidth,keepaspectratio]{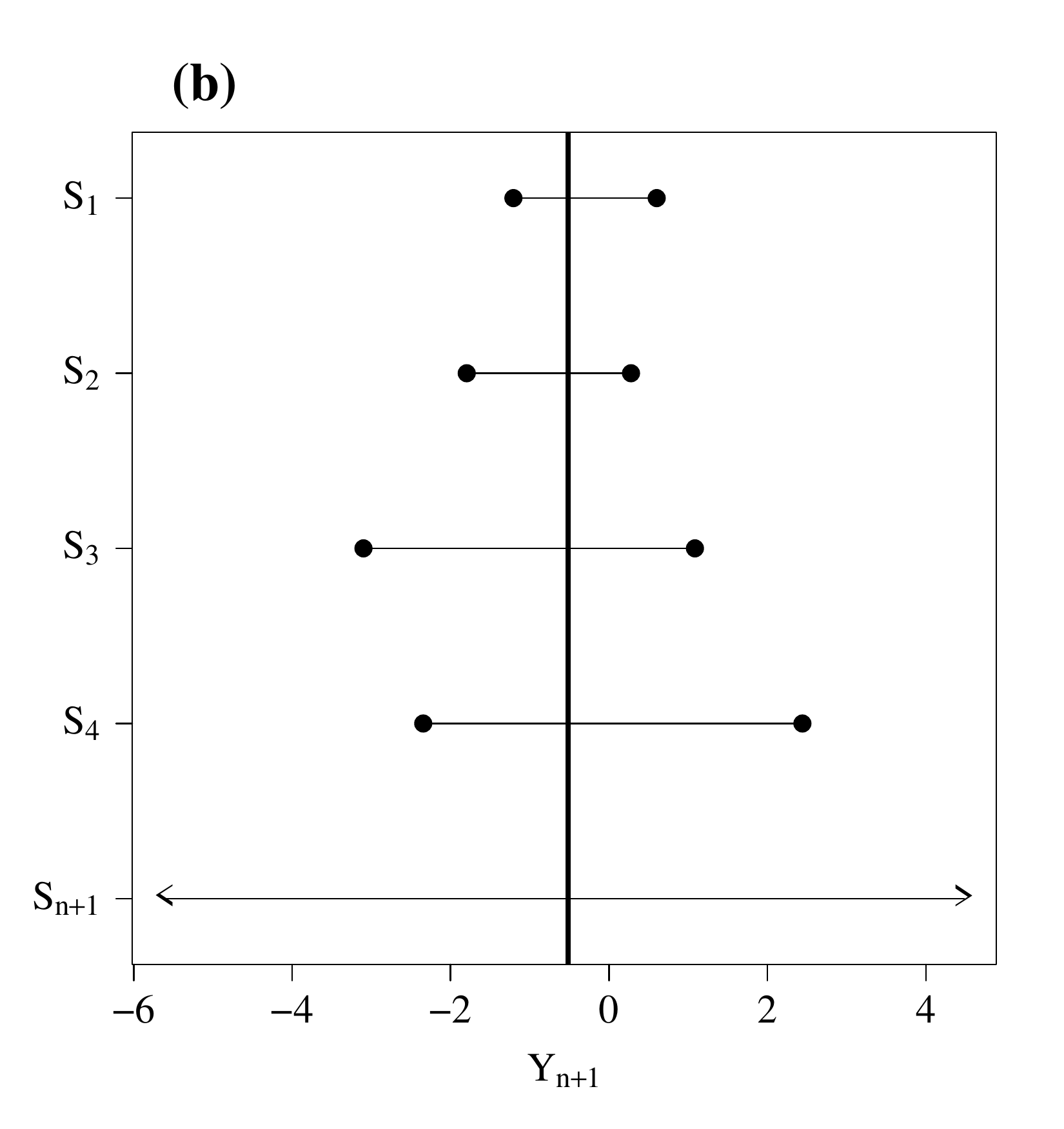} 
\includegraphics[width=.323\textwidth,keepaspectratio]{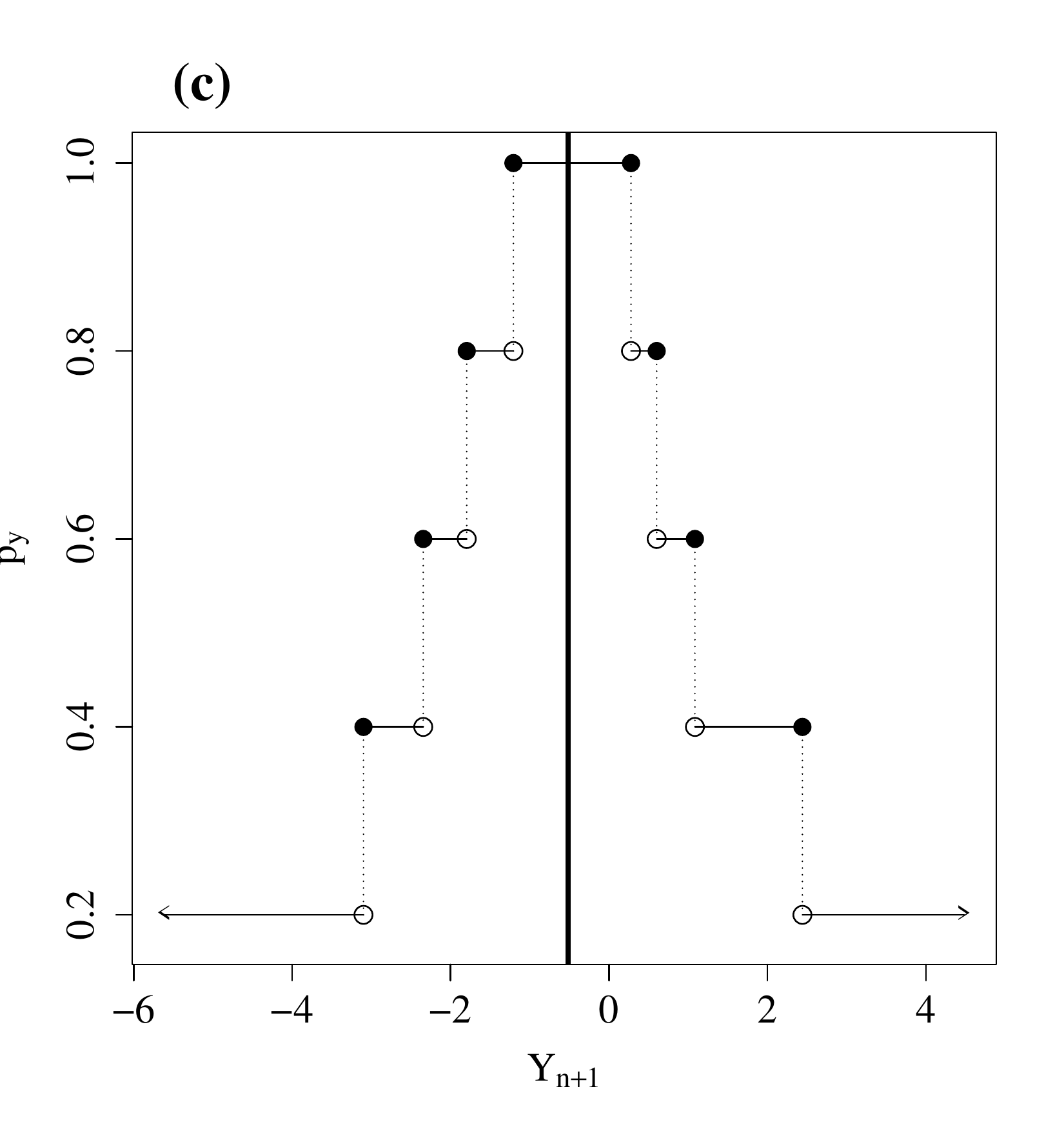}
\caption{Visualization of the process to obtain the FAB conformal region:
(a) conformity scores for each value in the sample (red dashed curves) and the candidate (thick black curve) over the sample space;
(b) sub-regions of acceptance corresponding to the conformity scores;
(c) number of conformity scores less than or equal to the candidate conformity score over the sample space;
vertical black lines are drawn at $\tilde{\theta}$.
}\label{compex}
\end{figure}

This process is visualized for an example sample of size $n=4$ in Figure \ref{compex}.
For clarity, the conformity scores for each value in the sample and the candidate prediction are plotted as a function of the candidate prediction in panel (a). 
The corresponding sub-regions of acceptance 
are the regions in the sample space where each conformity score is less than or equal to the conformity score of the candidate.
Under the normal working model, each sub-region of acceptance, plotted in panel (b), is an interval that contains $\tilde{\theta}$.
This information can be directly translated to the number of conformity scores less than or equal to the candidate conformity score over the sample space. Dividing this value by $(n+1)$ yields the conformal $p$-value, $p_y$ (shown in panel (c)).
From Figure \ref{compex}(c), it is easy to see for a prediction error rate of, for example, $\alpha = 0.2$, the resulting conformal prediction region is the region where $\#(i:c_i\leq c_{n+1})>1$, which is 
$[-3.1,2.4]$.

Given the standard form of the Bayes-optimal conformity measure (Equation \ref{postpred}), the sub-regions of acceptance are difficult to obtain analytically as the candidate $y_{n+1}$ appears non-linearly in each $c_{B,1}(y_{n+1}),...,c_{B,n+1}(y_{n+1})$.  
Upon consideration of the equivalent representation of $C_B$ given in Theorem \ref{conformitytheorem}, the regions $S_1,...,S_n$ can be expressed in closed form. This allows for efficient and exact computation of the prediction region and, in turn, can be used to prove the FAB prediction region is an interval under the normal working model.
These results are formalized below.

We first present two general lemmas (\ref{stepwise} and \ref{interval}) used to prove a conformal prediction region
is an interval.  
If the conformal $p$-value 
is a step function over the domain $\mathbb{R}$ with 
a symmetric number of unit steps to and from $1/(n+1)$ and $1$, 
as in Figure \ref{compex}(c),
then the prediction region will be an interval. 
The following two lemmas may be used to prove this is the case.

\begin{lemma}\label{stepwise}
Under the conformal algorithm, if for each $i=1,...,n+1$, the region
\[
\{y_{n+1}\in\mathbb{R}:c_i(y_{n+1})\leq c_{n+1}(y_{n+1})\}
\]
is an interval and contains some common value, then $f(y_{n+1})=\#\{i\in\{1,...,n+1\}:c_i\leq c_{n+1}\}$ is a step function 
with ordered output $1,2,...,n,n+1,n,...,2,1$ over the domain $\mathbb{R}$.
\end{lemma}

\begin{lemma}\label{interval}
Under the conformal algorithm, if $f(y_{n+1})$ is a step function 
with ordered output $1,2,...,n,n+1,n,...,2,1$ over the domain $\mathbb{R}$,
then the resulting prediction region is an interval.
\end{lemma}

It turns out the hypothesis of Lemma \ref{stepwise} holds for the normal working model case under $C_B$.
Specifically, by utilizing the equivalent form of $C_B$ given by Theorem \ref{conformitytheorem}, we are able to conclude that each $S_i$ is an interval for $i\in\{1,...,n\}$ and obtain a closed form expression of the bounds:

\begin{lemma}\label{smallinterval}
Under Model \ref{workingmodel}, for each $i=1,...,n$, the region
\[
S_i =\{y_{n+1}\in\mathcal{Y}:C_{B}\left(\{\boldsymbol y,y_{n+1}\}\backslash y_{i},y_{i} \right)\leq C_{B}\left(\{\boldsymbol y,y_{n+1}\}\backslash y_{n+1},y_{n+1} \right)\},
\]
is an interval $\left[\min\{y_i,g(y_i)\},\max\{y_i,g(y_i)\}\right]$ where
\[
g(y_i):= \frac{2\left(\mu/\tau^2+\sum_{k\in\{1:n\}} y_k \right)(1/\tau^2+n+1)^{-1} - y_i }{1-2(1/\tau^2+n+1)^{-1}}.
\]
\end{lemma}

Additionally, the posterior mean estimator of the population mean, $\tilde{\theta}$, is contained in each sub-region of acceptance $S_1,...,S_{n+1}$.
Not only is this result useful for proving the prediction region is an interval, 
but it also suggests that using $\tilde{\theta}$ as the estimator for a new prediction 
and 
taking $A^{fab}(\Ybf)$ as the prediction interval is a coherent method to predict the unknown value in a classically Bayesian manner while providing uncertainty quantification that maintains the desired frequentist coverage rate:

\begin{lemma}\label{containsshrinkage}
For each $i=1,...,n+1$, the interval 
\[
S_i =\{y_{n+1}\in\mathcal{Y}:C_{B}\left(\{\boldsymbol y,y_{n+1}\}\backslash y_{i},y_{i} \right)\leq C_{B}\left(\{\boldsymbol y,y_{n+1}\}\backslash y_{n+1},y_{n+1} \right)\},
\]
 contains the posterior mean estimator of the population mean, $\tilde{\theta} :=(\mu/\tau^2 + \sum_{k=1}^n y_k)(1/\tau^2+n)^{-1}$.
\end{lemma}

In total, these results 
can be used to
prove our main theorem concerning
properties of the FAB conformal prediction region constructed under the normal working model:

\begin{theorem}\label{maintheorem}
Under Model \ref{workingmodel} and conformity measure $C_{B}$, the 
$\alpha$-level conformal prediction region based on a sample $\Ybf$ is the $k$th and $(2n-k+1)$th order statistic of $\boldsymbol v$:
\[
A^{fab}(\Ybf) = \left( \boldsymbol v_{(k)}, \boldsymbol v_{(2n-k+1)}\right)
\]
for $ \boldsymbol v = \begin{bmatrix}y_1&\cdots&y_n&g(y_1)&\cdots&g(y_n)\end{bmatrix}^T$ and $k=\lfloor\alpha(n+1)\rfloor$.
Furthermore,
the conformal prediction region is an interval which contains $\tilde{\theta}$.
\end{theorem}

The nature of $C_B$ under the normal working model
suggests a simple, efficient algorithm to obtain $A^{fab}(\Ybf)$.
In particular, 
as $p_y$ is an incremental step function over the sample space characterized by a symmetry in the number of steps on either side of $p_y = 1$ (e.g. Figure \ref{compex}(c)), 
the $1-k/(n+1)$ prediction region can be obtained by taking the $k$th ordered step location from either end of the collection of step locations, where $k=\lfloor\alpha(n+1)\rfloor$.
Specifically,
to obtain $A^{fab}(\Ybf)$,
\begin{enumerate}
    \item for each $i=1,...,n$, obtain $y_i,g(y_i)$, the two critical values of $S_i$;
    \item set  $ \boldsymbol v = \begin{bmatrix}y_1&\cdots&y_n&g(y_1)&\cdots&g(y_n)\end{bmatrix}^T$ and $k=\lfloor\alpha(n+1)\rfloor$;
    \item acquire the 
    bounds of the prediction region,
    the $k$th and $(2n-k+1)$th order statistics of $\boldsymbol v$.
\end{enumerate}
Then, the Bayes-optimal conformal prediction region with $(1-k/(n+1))100\%$ coverage is
\[
A^{fab}(\Ybf) = \left( \boldsymbol v_{(k)}, \boldsymbol v_{(2n-k+1)}\right).
\]
If $\alpha (n+1)=k $, then the prediction region will have exact coverage.
A brief note that if $y_i=g(y_i)$ for at least one $i\in\{1,...,n\}$, the resulting prediction region
may be a point, dependent on the specified error rate.

\section{Numerical Comparisons}\label{singlepopsim}

\begin{figure}[h]
\centering
\includegraphics[width=.5\textwidth]{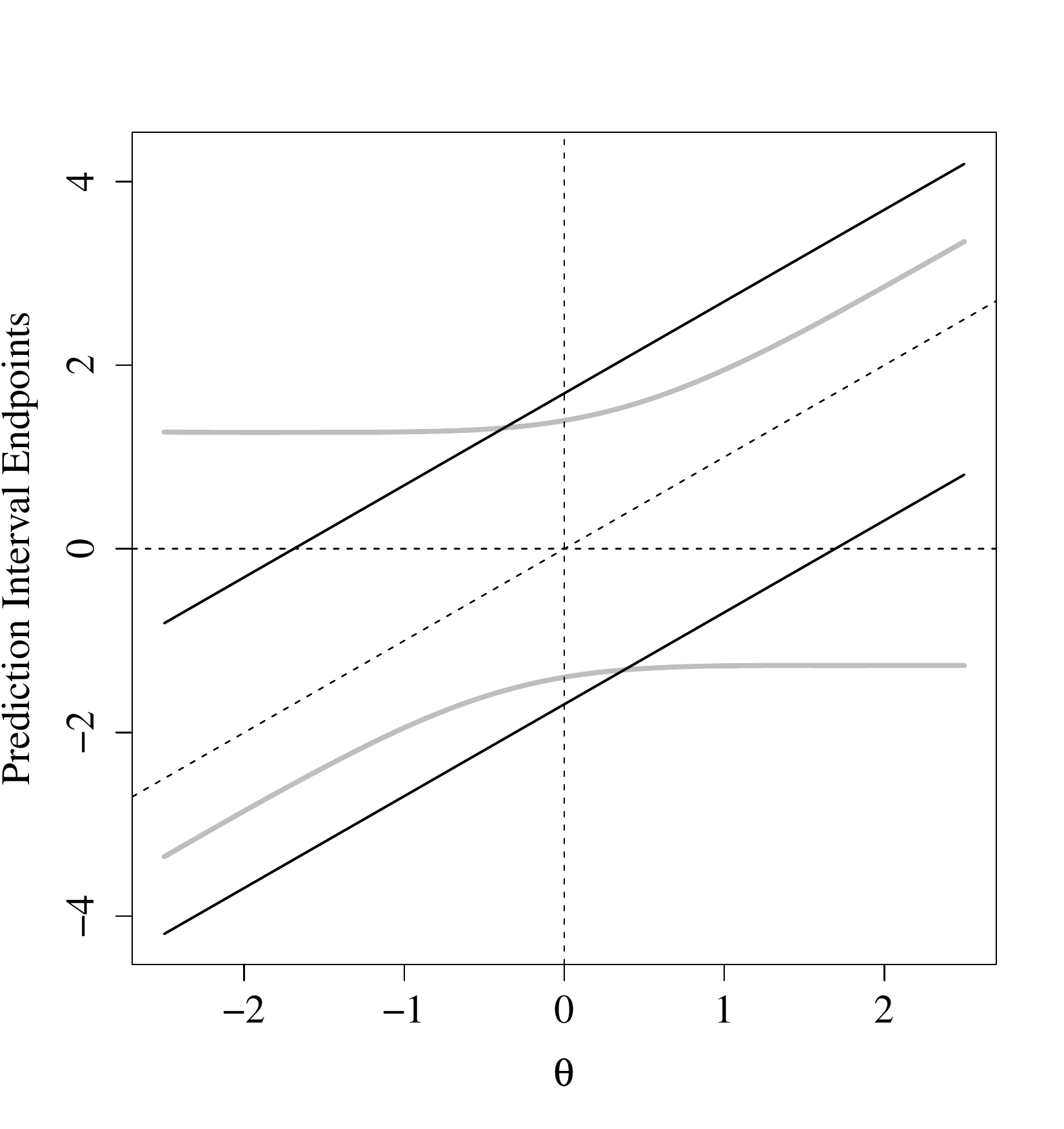}
\caption{Expected FAB (thick grey lines)
and DTA (solid black line)
prediction interval end points.
For $n=3,\theta\in[-2.5,2.5],\mu=0,\tau^2=1/2$.}\label{bounds}
\end{figure}

To demonstrate properties of the FAB prediction method,
we numerically evaluate expected prediction interval widths 
of the FAB prediction regions
and compare them to a simpler conformal method that does not use indirect information,
the distance to average (DTA) conformal prediction method. 
Both methods provide nonparametric frequentist coverage guarantees;
the main difference between these two methods is the ability to utilize prior information in the construction of the prediction interval.
While these prediction methods may be applied to non-normal populations, in this section,
we explore their behavior
when obtained from
a sample of size $n$
independently drawn from a normal population 
with mean $\theta$ and variance 1. 
As shown in Figure \ref{bounds}, the FAB method shifts prediction bounds away from the true population mean and towards the incorporated prior information. 
When this prior information well-informs the true mean, the FAB interval is narrower than the alternative. As the accuracy of the prior information declines, the FAB intervals must widen in order to maintain the desired coverage rate.
The risk and Bayes risk of the FAB conformal and distance to average conformal prediction intervals, where the risk function is taken to be expected interval width, $E_\theta[|A(\Ybf)|]$, is explored further in this section.

The prior parameters used in the FAB conformal method are $\{\mu,\tau^2\}$ and respectively represent the 
prior expected population mean and confidence in this expectation.
To assess the effect of these prior parameters,
we allow them to vary in this study. 
Additionally, we consider a prediction error rate of $0.25$ and
compare numerical results for various sample sizes $n\in\{3,7,11,15,19\}$,
chosen such that the conformal methods will result in regions with exact coverage.
In general, the FAB conformal method
outperforms 
the standard DTA method 
in terms of precision 
when 
there is concentrated and accurate prior information regarding the mean of the population, but a limited amount of information in the sample itself.
More specifically, the FAB conformal interval can be expected to produce narrower intervals than standard methods when $|\theta-\mu|$ is small, $\tau^2$ is small, and $n$ is small, or a combination of these properties. 



\begin{figure}[h]
\centering
\includegraphics[width=.323\textwidth]{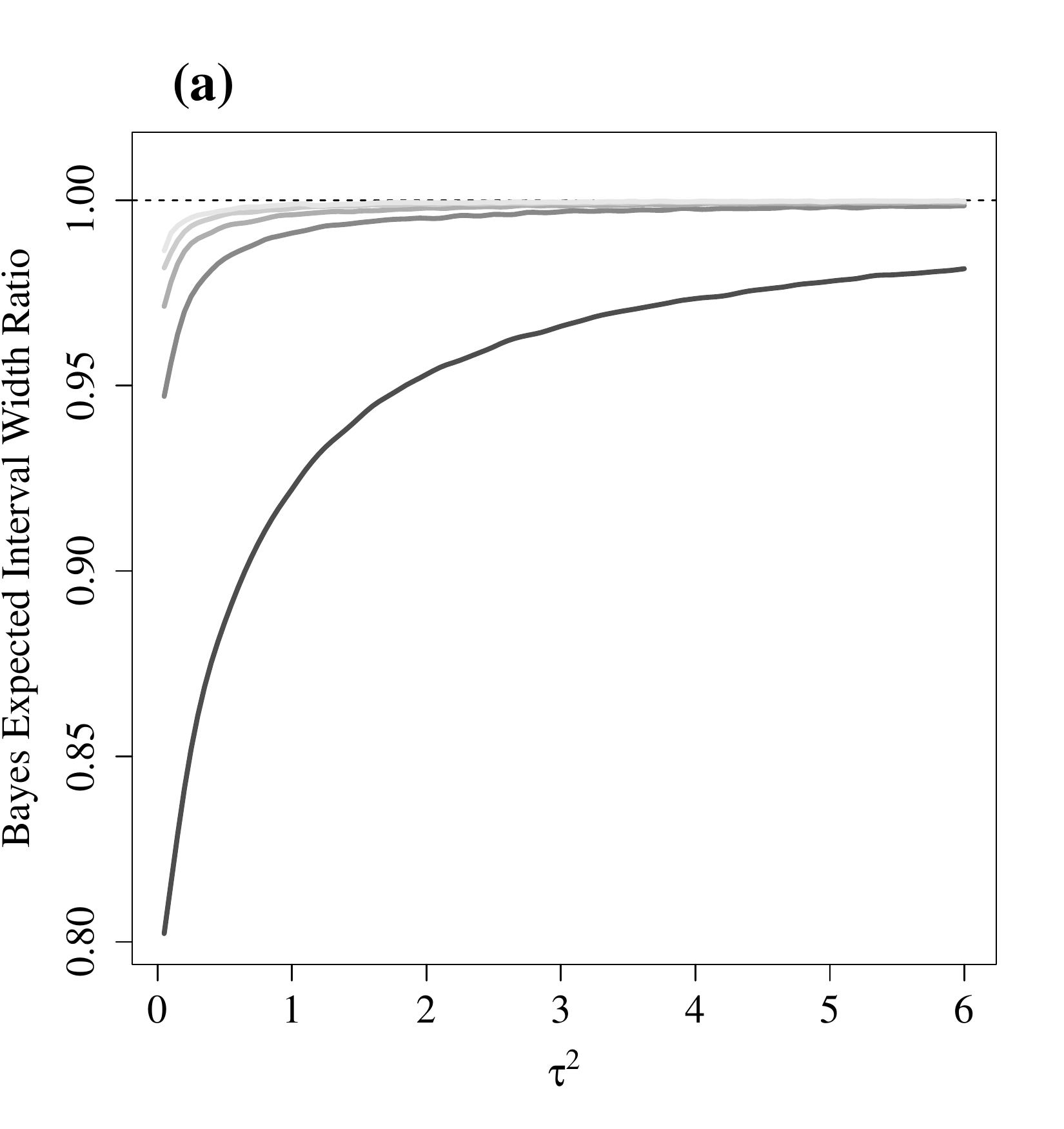}
\includegraphics[width=.323\textwidth]{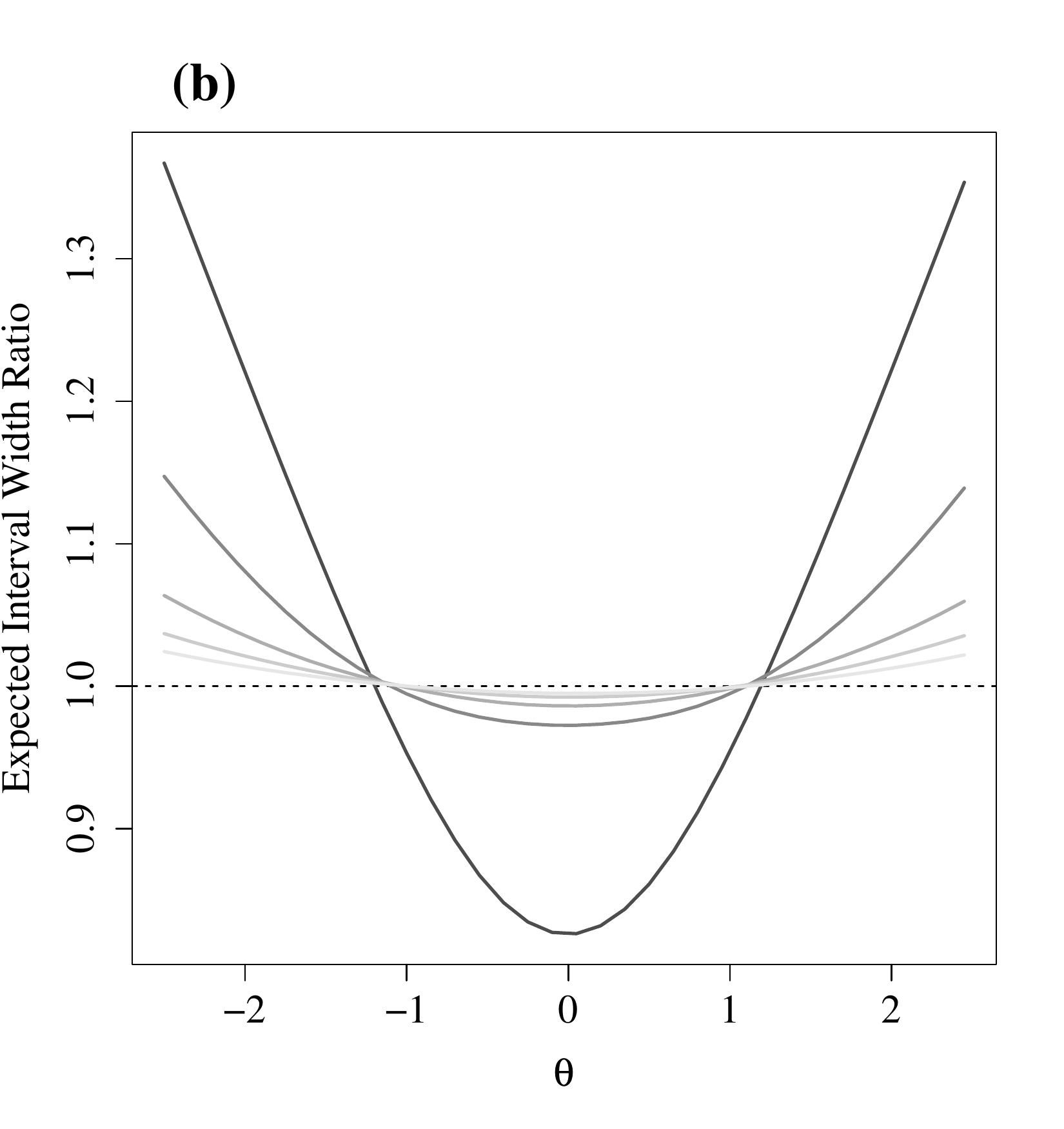}
\includegraphics[width=.323\textwidth]{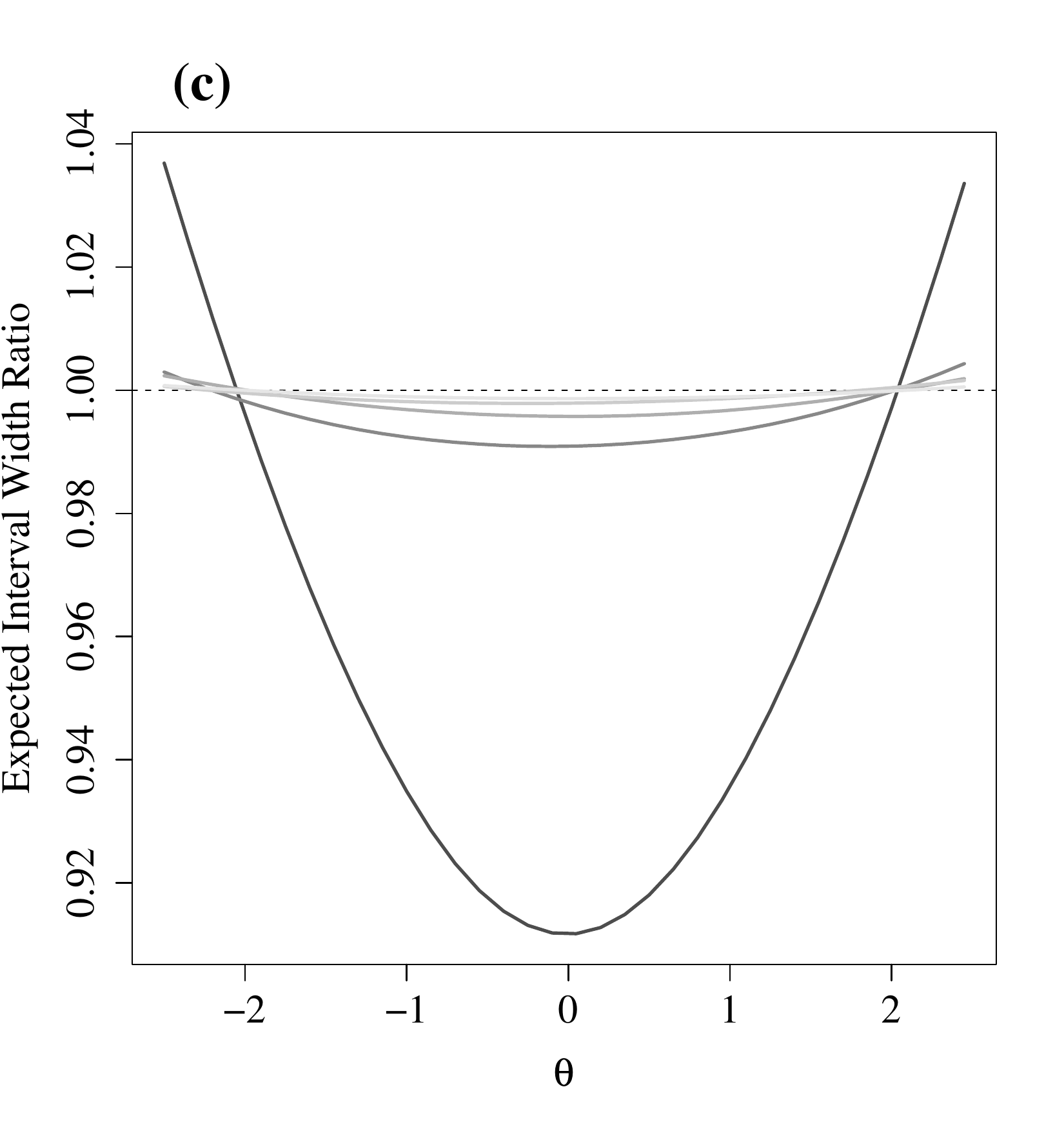}
\caption{Expected width ratio of 
FAB conformal interval to distance to average conformal interval for increasing $n\in\{3,7,11,15,19\}$ in decreasing darkness:
(a) expectation taken with respect to $\Ybf$ and $\theta$;
(b) expectation taken over $\Ybf$ conditional on $\theta$ for  $\mu=0,\tau^2=1/2$;
(c) same as (b) for $\tau^2=2$.}\label{single_pop}
\end{figure}


The ratio of expected interval widths of the FAB prediction method relative to the distance to average method are displayed in Figure \ref{single_pop}. Recall that the FAB method incorporates prior information while the distance to average method does not. We compute the expected interval widths via Monte Carlo 
approximation using 25,000 independently generated replications for each combination
of values of $\theta,\tau^2,$ and $n$.
The effects of sample size and prior variance of the population mean are
the focus of Figure \ref{single_pop}(a).
This figure plots
the ratio of 
Bayes expected interval widths of the FAB conformal to distance to average conformal intervals,
where the expectations are taken with respect to $\Ybf$ and $\theta$ under 
the sampling model $Y_1,...,Y_n\sim N(\theta,1)$ and 
prior $\theta \sim N(\mu,\tau^2)$. 
The Bayes risks of the conformal intervals do not depend on $\mu$. 
As informed by the theoretical Bayes-optimality of FAB conformal prediction,
the Bayes risk of the FAB interval is smaller than that of the DTA interval,
with
the overall deviation between the methods' expected interval widths decreasing as the sample size increases.
The FAB interval is substantially narrower than the distance to average interval for a wide range of $\tau^2$ values under very small sample sizes.
Intuitively, for small sample sizes, even a low level of confidence around the prior value of the population mean $\mu$ is useful information and will translate to narrower prediction intervals if utilized in the construction of $C_B$.
More confidence, as conveyed through a smaller $\tau^2$ value, translates to a more substantive increase in precision.
Even for larger sample sizes, a nontrivial gain in precision occurs under small (less than about 0.5) $\tau^2$ values, 
representing very informative prior information about the population mean through a prior with tight concentration around $\mu$. 

For a given concentration level $\tau^2$,
it is natural to consider 
how the accuracy of the mean estimate $\mu$ affects the resulting FAB prediction interval width.
Figures \ref{single_pop}(b)-(c) display the expected interval width ratio of the FAB conformal to the distance to average conformal method for varying population means 
and sample sizes when
$\mu = 0$ and $\tau^2=1/2,2$.
Under this set-up, by the Bayes-optimal property of the conformal measure, the FAB interval will
outperform alternatives when $|\theta-\mu|\approx 0$ and $\tau^2$ is small. 
The numerical results match this conclusion, and, as the distance between the 
prior mean and the population mean increases in absolute value, the FAB intervals become wider, and thus the benefit of utilizing this type of prior information declines.
As seen in panel (b), for a moderately small $\tau^2$ value, and for the smallest sample size considered, the FAB conformal method results in an interval width that is 17.6\% narrower than the standard when $\theta=\mu$ exactly. 
For larger $\tau^2$, as seen in the panel (c), there is a less substantial benefit to utilizing this auxiliary information in $C_B$, but some benefit is seen nonetheless for a wider range of $\theta$ divergences from $\mu$.

Overall, when
accurate prior information is available,
FAB prediction intervals outperform commonly utilized prediction intervals as judged by precision and coverage guarantees, and thus
there will be a benefit to exploiting this information in obtaining prediction intervals via the FAB method. 
The benefit is particularly large when obtaining prediction intervals based on small sample sizes.
Regardless of the accuracy of the prior information, the frequentist coverage rate of FAB prediction intervals 
is guaranteed.


\section{FAB Small Area Prediction}\label{fabsapmain}

\subsection{Information Sharing via a Working Model}

In inference on small areas, utilizing indirect methods that share information across areas has been shown to improve precision compared with direct methods, 
particularly for areas with small sample sizes (\cite{Gelman2006}). 
With this motivation, we extend the FAB conformal prediction method 
explored thus far to 
a small area regime.
In the construction of the Bayes-optimal conformity measure,
information is shared across groups via a
multilevel working model 
in order to increase prediction region precision
while maintaining area-level frequentist coverage guarantees.

For group $j\in\{1,...,J\}$, we observe an exchangeable sample $(Y_{1,j},...,Y_{n_j,j})=(y_{1,j},...,y_{n_j,j})$ of length $n_j$ such that the 
samples are independent across groups.
Suppose a reasonable working model for the populations is a spatial Fay-Herriot model (\cite{Fay1979}) that allows for heterogeneous area-specific variances.
Specifically, 
\begin{align}\label{workingmod}
Y_{1,j},...,Y_{n_j,j}\sim{}& N(\theta_j,\variance_j),\; \text{independently for $j=1,...,J$}\\
\thetabf\sim{}&N\left(\boldsymbol X\boldsymbol\beta,\eta^2\boldsymbol G \right)\nonumber \\
\variance_1,...,\variance_J\sim{}&IG(a/2,b/2),\nonumber
\end{align}
where $\boldsymbol G$ is a spatial covariance matrix such as that which results from the popular simultaneous (SAR) autoregressive model
$\boldsymbol G =  [(\boldsymbol I-\rho \boldsymbol W)(\boldsymbol I-\rho \boldsymbol W^T)]^{-1}$ (\cite{Singh2005}).
The matrix $\boldsymbol W$ is a distance matrix among areas that is typically row-standardized to sum to 1, and $\rho\in(-1,1)$ is a spatial correlation parameter. For more on spatial modeling, see \cite{Banerjee2014}.
This flexible set-up allows for inclusion of an array of auxiliary information including area-level covariates $\boldsymbol X$ and spatial relationships in the linking model for the population means which can be exploited to improve precision of prediction regions.

For population $j$, 
we may estimate unknown parameters in the working model (Eqns \ref{workingmod}) given all available data outside of area $j$,
$\Ybf_{-j}:=\{\Ybf_1,...,\Ybf_J\}\backslash\{\Ybf_j\}$,
and use this information to inform the unknown prior parameters in the Bayes optimal conformity measure.
A FAB conformal predictive interval may then be constructed given this information.
Specifically, given prior parameter values and 
the independence of $\Ybf_1,...,\Ybf_J$, 
the form of the Bayes optimal conformity measure for population $j$ is as given in Equation \ref{postpred}. 
The resulting prediction region is constructed from a measure that shares information across areas and will maintain the desired frequentist coverage rate for each area.

\subsection{FAB Conformal Parameter Estimation Procedure}\label{estprocedure}

All that remains 
to proceed with the FAB conformal method
is to acquire values for the unknown prior parameters needed for the Bayes-optimal conformity score.
To apply the method to area $j$,
we require estimates of
$\mu$, the prior mean of area $j$'s population mean $\theta_j$,
and $\tau^2$, the ratio of the prior variance of $\theta_j$ to population $j$'s variance $\variance_j$,
obtained from data in other groups. 
If the parameters of the working model are known, we can 
take $\{\mu,\tau^2\}$ to be
the conditional mean and conditional variance proportion of $\theta_j$ and proceed with implementation of the Bayes-optimal conformal algorithm. 
In practice, of course, these values are not known, but they may be estimated via standard techniques.

We propose an empirical Bayesian approach whereby
values of the prior parameters
$\{\mu,\tau^2\}$ are obtained for each group $j$ using samples from all other populations, $\Ybf_{-j}$.
As an overview, for area $j$, our estimation procedure 
proceeds with 
first computing estimates of unknown parameters in the working model using $\Ybf_{-j}$. 
Then, given these estimates, we obtain
the conditional mean of $\theta_{j}$ and
the proportion of the conditional variance of $\theta_{j}$ to an estimate of area $j$'s population variance,
which are labeled as $\mu_j$ and $\tau^2_j$, respectively.

In more detail, first obtain the maximum likelihood estimates (MLEs) of $a,b$ by maximizing the marginal density of 
$\{S_k^2\}_{k\in K}$ 
for $K=\{1,...,J\}\backslash j$
where  $S^2_k=\sum_{i=1}^{n_k}(\Ybf_{ik}-\bar{\Ybf}_k)$ . 
Under the assumptions of the working model (\ref{workingmod}),
the marginal density for the entire population can be shown to be
\[
p(s_1^2,...,s_J^2|a,b) = \prod_{j=1}^Jf(s_j^2)\frac{\Gamma\left(\frac{a +n_j-1}{2}\right)
\left(\frac{b}{2}\right)^{a/2}}
{\Gamma \left(\frac{a}{2}\right)\left(\frac{b+s^2_j}{2}\right)^{(a+n_j-1)/2}}
\]
for a function $f$ that does not depend on the hyperparameters $a,b$.
We use the resulting MLEs to obtain empirical Bayes estimates of each area's population variance.
That is, take
$\hat{\sigma}^2_k = (\hat{b}+s_k^2)(\hat{a}+(n_k-1)+1)$ for $k\in\{1,....K\}$
and 
$\hat{\sigma}_{j}^2 = \hat{b}/(\hat{a}+1)$.
Taking $\{\hat{\sigma}_k^2\}_{k\in K}$ as plug-in values of the population variances,
we obtain the marginal maximum likelihood estimates of the mean prior hyperparameters $\boldsymbol\beta, \rho$, and $\eta^2$ through standard REML or ML procedures (\cite{Pratesi2008}).
Maximum likelihood estimates $\{\hat{\boldsymbol\beta},\hat{\eta}^2,\hat{\rho}\}$ of $\{\boldsymbol\beta,\eta^2,\rho\}$ may then be used to obtain an empirical Bayes estimate $\hat{\boldsymbol \theta}$ of $\boldsymbol \theta$.

For area $j$, we obtain estimates 
$\{\hat{\boldsymbol\beta},\hat{\eta}^2,\hat{\rho},\hat{\boldsymbol \theta},\{\hat{\sigma}_k^2\}_{k\in K},\hat{\sigma}_{j}^2\}$ 
as specified above using
$\Ybf_{-j}$. 
Then, 
the prior parameters of area $j$ are the conditional mean and the proportion of conditional variance of $\theta_j$ obtained under these estimates:
\begin{align}
    {\mu}_j ={}& E[\theta_j|\boldsymbol\theta_{-j}]\\
    ={}& \boldsymbol{x}_j^T  \hat{\boldsymbol\beta} + \boldsymbol V_{[j,-j]} \boldsymbol V_{[-j,-j]}^{-1}\left(\hat{\boldsymbol \theta}_{-j} - \boldsymbol X_{[-j]}\hat{\boldsymbol\beta}\right)\nonumber\\
    {\tau}_j^2 ={}& Var[{\theta}_j|{\boldsymbol\theta}_{-j}]/\hat{\sigma}^2_j\\
    ={}& \left({\boldsymbol V}_{[j,j]} - 
    {\boldsymbol V}_{[j,-j]}{\boldsymbol V}_{[-j,-j]}^{-1}{\boldsymbol V}_{[-j,j]}\right)/\hat{\sigma}^2_j\nonumber
\end{align}
where $\boldsymbol V = \hat{\eta}^2[(\boldsymbol I-\hat{\rho} \boldsymbol W)(\boldsymbol I-\hat{\rho} \boldsymbol W^T)]^{-1}$.
Given these values for $\mu$ and $\tau^2$, obtained from information independent of area $j$, the
conformal algorithm proceeds as described in Section \ref{fabcomputation}.
For each area, the algorithm yields an interval that may have improved precision over other methods as a result of information sharing and maintains the specified frequentist coverage rate.

\section{Radon Data Example}\label{radonexample}

Indoor radon levels are a significant risk factor to public health. 
To better understand the risk, the U.S. Environmental Protection Agency (EPA) conducted a national survey (\cite{USEnvironmentalProtectionAgency1992})
on indoor radon levels.
They gathered indoor radon readings from a stratified random sample of households across 9 states.  
We limit our scope to a subset of the available data and explore the sample collected in Minnesota
which consists of 919 observations in total throughout the state's 85 counties. 
While within-county sample sizes range from 1 to 116, many are quite small. In particular,
46\% of within-county sample sizes are 5 or less.

\cite{Price1996} explored modeling the household radon levels in Minnesota with a goal of improving estimated county-level means, and accurate county-specific predictions are cited as being of particular interest. Due to the frequency of small within-county sample sizes, these are difficult tasks.  
Given the abundance of auxiliary information, including county-level covariates and apparent spatial relationships among radon levels across counties, 
incorporating indirect information
in the construction of confidence or prediction intervals is a natural tool to improve inferential precision. 
In what follows, we compare properties of prediction intervals resulting from FAB and DTA methods.
As county-specific predictive inference is of primary interest, an ideal prediction interval will have the desired $(1-\alpha)100\%$ frequentist coverage for every county while maintaining an interval width that is practically informative.

We construct prediction intervals
for log radon at one new randomly sampled household in each county
where $Y_{ij}$ is the log radon value for household $i$ in county $j$. 
We assume that $Y_{ij}$ is independently distributed within and across counties.
Exploratory data analysis suggests log radon values are approximately normally distributed, so we utilize the normal working model (Equations \ref{workingmod}) in the construction of the FAB conformal prediction intervals.
For each county $j\in\{1,...,J\}$,
estimates of the prior parameters are obtained using data from all counties outside of $j$, as described in Section \ref{estprocedure}. 
We include as a covariate a county-level soil uranium measurement, incorporate a shared county-level prior intercept, and allow for spatial effects under the (row-standardized) squared exponential distance matrix $\boldsymbol W$ between county centroids. That is, before row-standardization, the matrix entries are $\{w_{lk}\} = e^{- ||x_l-x_k||^2}$ for counties $l,k$.

\begin{figure}[h]
\centering
\includegraphics[width=.95\textwidth,keepaspectratio]{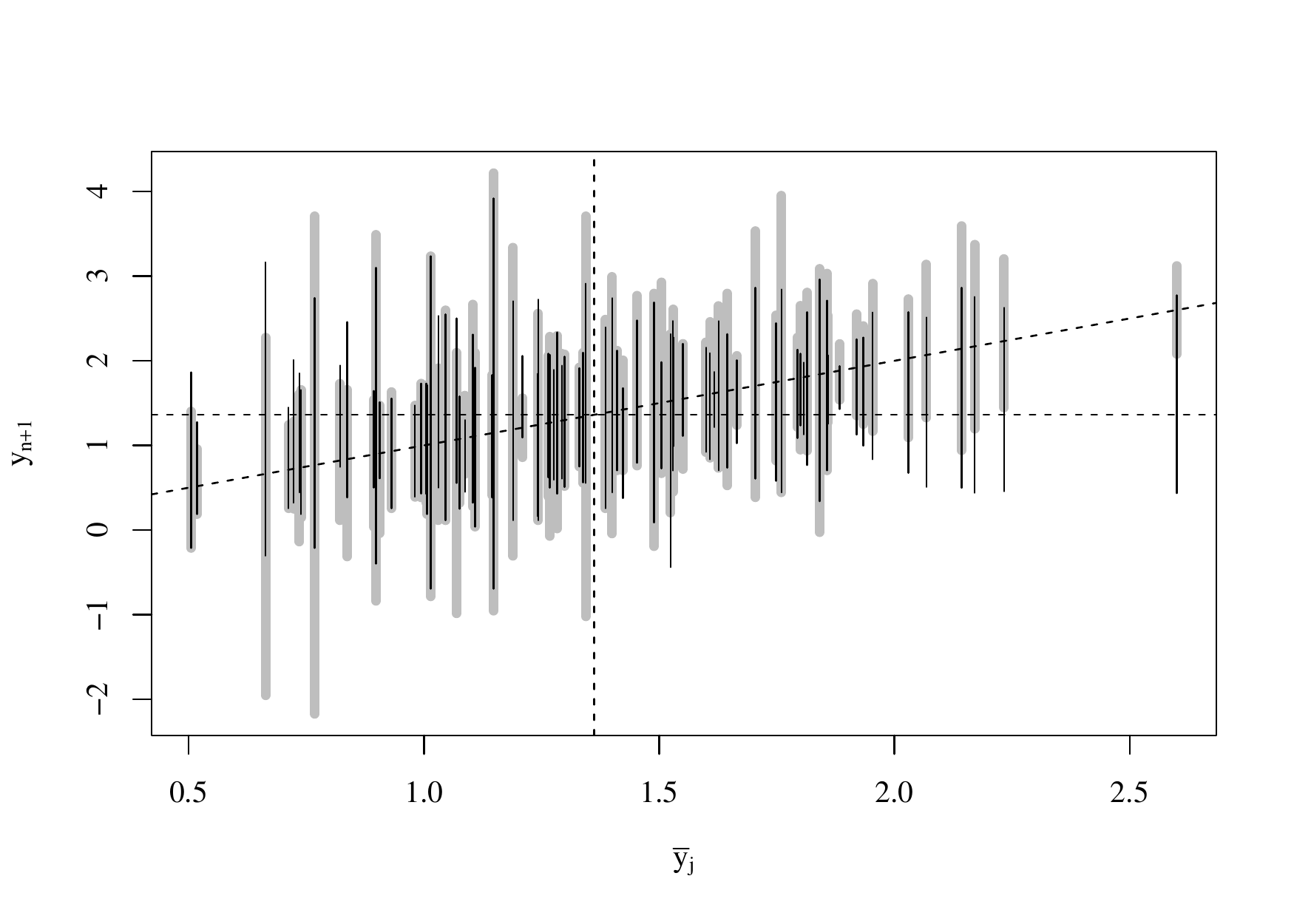}
\caption{County-level radon prediction intervals: 
black solid lines represent the FAB conformal intervals and thick grey lines represent the distance to average conformal intervals.
Dashed lines drawn at the state-wide sample mean $\sum \bar{y}_j/J$ and 45\degree\hspace{1mm} line.}\label{radon_predints}
\end{figure}

FAB conformal and DTA conformal prediction intervals are obtained for counties with sample sizes greater than 1
under a county-specific error rate $\alpha_j = \lfloor\frac{1}{3}(n_j+1)\rfloor/(n_j+1)$ to allow for
exact $1-\alpha_j$ frequentist coverage of prediction intervals for each county $j$.
The prediction intervals for each county are plotted in Figure \ref{radon_predints}. 
In summary, including relevant auxiliary information in the construction of prediction intervals via the Bayes-optimal conformal procedure results in improved overall interval precision.
Specifically, the FAB prediction intervals are narrower than the DTA intervals in 68.3$\%$ of counties.

The FAB intervals exhibit classical `Bayesian' or shrinkage behavior in that they are shifted towards the shared state-wide sample mean, while each DTA interval is centered near the respective county-specific sample mean.
By the Bayes-optimality property of the FAB prediction method, FAB prediction intervals are narrower than all alternative prediction regions (for a specified prediction error rate) for counties
where the working model assumptions 
are accurate.
As such, in this case, the FAB intervals perform best for counties where the heterogeneity across county-specific mean radon values is well described by the provided spatial and covariate information.
As shown in Figure \ref{radon_predints}, the FAB prediction intervals are frequently narrower than the DTA intervals even though they maintain the same coverage rates.
There are, however, some counties where the FAB intervals are wider than the DTA intervals. 
As shown in the numerical results, this behavior is expected in cases where the utilized prior information is not accurate,
which practically corresponds to outlier counties
where available indirect information does not inform that counties population-specific mean.
County-specific precision is sacrificed in these counties.
In conclusion, sharing information via the FAB conformal method
can result in narrower prediction intervals than standard methods.

\section{Discussion}\label{concl}

When constructed under a normal working model,
FAB conformal prediction regions are intervals that contain the standard posterior mean estimator $\tilde{\theta}$.
This implies a coherent method of Bayesian point prediction, through $\tilde{\theta}$,
where uncertainty quantification maintains the specified frequentist coverage rate.
Furthermore, exact FAB conformal prediction regions may be obtained efficiently.
The FAB conformal prediction method introduced in this article,
which utilizes the posterior predictive density of some working model
as a conformity measure,
produces precise and accurate prediction regions.
If the working model assumptions are accurate, 
FAB prediction regions have minimum expected volume, and
regardless of the accuracy of the working model, 
they have guaranteed frequentist coverage rate control.

In practice, FAB conformal intervals are notably narrower than standard prediction intervals in the case that accurate prior information regarding the population mean is available with a high degree of confidence, especially
in the case of small sample sizes. 
In short, the FAB conformal prediction method
leverages this accurate indirect information to improve interval width precision.
As such,
this method is particularly useful for small area applications
where area-specific frequentist coverage guarantees are desirable and sample sizes are often prohibitively small. 
While commonly used small area prediction methods involve a trade-off between accuracy and precision,
the FAB prediction method maintains a guarantee of accuracy while allowing for increased precision via cross-area information sharing.


This article
focuses on FAB prediction intervals constructed under a normal working model, but
the Bayes-optimal conformity measure can be constructed under alternative working models that may be more appropriate for some types of data.
As such, we have presented a framework that may be utilized to aid in derivation of more efficient computation of conformal prediction regions.

Replication code and data are available at \url{https://github.com/betsybersson/fab_sap}.

\newpage
\typeout{}
\bibliography{./MiscFiles/library}
\bibliographystyle{./MiscFiles/spbasic}

\appendix

\section{Notation Simplification}

We begin by simplifying notation to be used in all proofs hereafter. 
In working with the Bayes-optimal conformity measure, we often find ourselves 
making comparisons between densities of two random variables conditional on each other and a shared set of random variables. 
The information contained in the shared set may be viewed as prior information, which results in fewer variables to keep track of.
Specifically,
for two random variables of interest, $Y_j,Y_k$, conditioning on extra data $\Ybf_{-j,-k}$ simply requires an update of the posterior of model parameters $\thetabf$.  
In terms specific to the conformal predictive procedure,
Lemma \ref{simplifylemma} allows us to consider a simplified regime where there is a single data point $y_j$ and a single candidate prediction $y_{n+1}$. 

\begin{lemma}\label{simplifylemma}
The density $p(Y_j|\{Y_1,...,Y_{n+1}\}\backslash Y_j)$  under an IID hierarchical model,
\begin{align}\label{simplifymodel1}
Y_1,...,Y_{n+1}\sim{}& P_\thetabf\\
\thetabf \sim{}& Q,\nonumber
\end{align}
is equivalent to $p(Y_j|Y_{k})$ under the prior density 
\[
\tilde{q}(\thetabf)=p_q(\thetabf|\Ybf_{-j,-k})= \frac{p_\theta(\Ybf_{-j,-k}|\thetabf)q(\thetabf)}{\int_\Theta p_\theta(\Ybf_{-j,-k}|\thetabf)q(\thetabf) d\thetabf}
\]
 where $\Ybf_{-j,-k} := \{Y_1,...,Y_{n+1}\}\backslash \{Y_j,Y_k\}$ for any $k\in\{1,...,n+1\},\;k\ne j$, and 
 $p_q(\thetabf|\cdot)$ refers to the posterior density of $\thetabf$ conditional on $(\cdot)$ under the prior $q$.
\end{lemma}

\begin{proof}[Proof of Lemma \ref{simplifylemma}]
We aim to show the distribution of $Y_j|\Ybf_{-j}$
under Model \ref{simplifymodel1} is equivalent to 
the distribution of $Y_j|Y_k$
under following model:
\begin{align}\label{simplifymodel2}
Y_j,Y_k\distiid{}& P_\thetabf\\
 \thetabf \sim{}&\tilde{Q}.\nonumber
\end{align}
First,
\begin{align*}
p_{\tilde{q}}(\thetabf|Y_k)\propto{}& p_\theta(Y_k|\thetabf)\tilde{q}(\thetabf) \\
 ={}& p_\theta(Y_k|\thetabf)p_q(\thetabf|\Ybf_{-j,-k})\\
\propto{}& p_\theta(Y_k|\thetabf)p_\theta(\Ybf_{-j,-k}|\thetabf)q(\thetabf)\\
={}& p_\theta(\Ybf_{-j}|\thetabf)q(\thetabf)\\
\propto{}& p_q(\thetabf|\Ybf_{-j})
\end{align*}
Therefore, $p_{\tilde{q}}(\thetabf|Y_k)\equiv p_q(\thetabf|\Ybf_{-j})$.

Then, under Model \ref{simplifymodel1}
\begin{align*}
p(Y_j|\Ybf_{-j})={}&\int _\Theta p_q(Y_j,\thetabf|\Ybf_{-j})d\thetabf\\
={}&\int _\Theta p_\theta(Y_j|\thetabf)p_q(\thetabf|\Ybf_{-j})d\thetabf\\
\equiv{}& \int_\Theta p_\theta(Y_j|\thetabf)p_{\tilde{q}}(\thetabf|Y_k)d\thetabf
\end{align*}
which is equivalent to $p(Y_j|Y_k)$ under Model \ref{simplifymodel2}.
\end{proof}

\section{Proofs}\label{proofsall}

One straightforward method for proving two measures are ECM is to show, for each $i=1,...,n+1$, the sub-region of acceptance, $S_i$,
$S_i = \{y_{n+1}\in\mathcal{Y}:c_i(y_{n+1})\leq c_{n+1}(y_{n+1})\}$,
is the same for both measures.
This generic result will be used in the proof of Theorem \ref{conformitytheorem}.

\begin{lemma}\label{CDequiv}
For conformity measures $C,D$, if
\[
\{y_{n+1}:c_i(y_{n+1})\leq c_{n+1}(y_{n+1})\}=\{y_{n+1}:d_i(y_{n+1})\leq d_{n+1}(y_{n+1})\} \;\forall\; i=1,...,n+1
\]
then $C$ and $D$ are ECM.
\end{lemma}

\begin{proof}[Proof of Lemma \ref{CDequiv}] 
\begin{align*}
\{y_{n+1}:{}&c_i\leq c_{n+1}\}=\{y_{n+1}:d_i\leq d_{n+1}\} \;\forall i=1,...,n+1\\
\Rightarrow {}&\#\{i:c_i\leq c_{n+1}\} = \#\{i:d_i\leq d_{n+1}\}\\
\Rightarrow {}&p_{y,c}=p_{y,d}
\end{align*}
where $p_{y,x}$ is the conformal $p$-value corresponding to conformity measure $x$.
Thus each candidate prediction value will be treated the same under both conformity measures $C,D$. 
\end{proof}

\begin{proof}[Proof of Lemma \ref{stepwise}] 
For each $i=1,...,n+1$, the region $S_i=\{y_{n+1}\in\reals:c_i(y_{n+1})\leq c_{n+1}(y_{n+1})\}$ is an interval and contains a shared value, labeled $\gamma$. 
Then,
$f(y_{n+1}|y_{n+1}<=\gamma)$ increases stepwise from 1 to $n+1$ over the range $(-\infty,\gamma]$.  
Likewise, $f(y_{n+1}|y_{n+1}>=\gamma)$ decreases stepwise from $n+1$ to 1 over the range $[\gamma,\infty)$.
Therefore, $f(y_{n+1})$ is a step function that takes on ordered values $\{1,2,...,n,n+1,n,...,2,1\}$
over the domain $\reals$.
\end{proof}

\begin{proof}[Proof of Lemma \ref{interval}] 
The region in $\reals$ where $f(y_{n+1}) >a$ for some $a\in [0,n+1]$ is an interval. Therefore,
\[
A(\Ybf) = \{y_{n+1}\in\mathcal{Y} :f(y_{n+1})/(n+1)>\alpha\}
\]
is an interval.
\end{proof}

\begin{proof}[Proof of Thm \ref{conformitytheorem}]
By Lemma \ref{simplifylemma}, we consider the conformity between two values $y_1,y_2$. 
Under the normal working model, in this case, the Bayes-optimal conformity measure is
\[
C_B\left(y_1,y_2\right) := p\left(y_2|y_1\right)=\frac{\Gamma \left(\frac{a_{12}+1}{2}\right)}{\sqrt{a_{12}\pi}\Gamma\left(\frac{a_{12}}{2}\right)} 
\left(\left(\frac{b_{2|1}}{a_{12}}(1+\tau^2_{12})\right)^{-1/2}
\left( 1  +\frac{1}{a_{12}}\frac{(y_2-\mu_{2|1})^2}{\frac{b_{2|1}}{a_{12}}(1+\tau^2_{12}) }\right)^{-(a_{12}+1)/2}
\right),
\]
where
\begin{align*}
{}& \tau^2_{12} = (1/\tau^2+1)^{-1}\\
{}&\mu_{2|1} = (\mu/\tau^2+y_1)\tau^2_{12}\\
{}&a_{12} = a+1\\
{}&b_{2|1} =b +y_1^2+\mu^2/\tau^2-(\mu/\tau^2+y_1)^2\tau^2_{12}.
\end{align*}

Now, suppose the conformal algorithm requires that we identify the region of $y_1$ s.t. $C_B\left(y_1,y_2\right) \leq C_B\left(y_2,y_1\right)$. This region can be shown to be the same as that obtained from $C_B\left(\{y_1,y_2\},y_2\right) \leq C_B \left(\{y_1,y_2\},y_1\right)$, which is the definition of equivalent conformity measures:
\begin{align}
{}&\hspace{7mm}C_B\left(y_1,y_2\right) \leq C_B \left(y_2,y_1\right) \nonumber\\
{}&\Leftrightarrow C_B\left(y_1,y_2\right)/C_B \left(y_2,y_1\right) \leq 1\nonumber\\
{}&\Leftrightarrow\frac{\left(b_{1|2}\right)
\left( 1  +\frac{(y_1-\mu_{1|2})^2}{b_{1|2}(1+\tau^2_{12}) }\right)^{a_{12}+1}}{
\left(b_{2|1}\right)
\left( 1  +\frac{(y_2-\mu_{2|1})^2}{b_{2|1}(1+\tau^2_{12}) }\right)^{a_{12}+1}}
\leq 1\nonumber\\
{}&\Leftrightarrow \frac{\left(b_{1|2}\right)
\left( \frac{b_{1|2}(1+\tau^2_{12}) +(y_1-\mu_{1|2})^2}{b_{1|2} }\right)^{a_{12}+1}}{
\left(b_{2|1}\right)
\left(\frac{b_{2|1}(1+\tau^2_{12})+(y_2-\mu_{2|1})^2}{b_{2|1} }\right)^{a_{12}+1}}
\leq 1\nonumber\\
{}&\Leftrightarrow  
\left(\frac{b_{2|1}}{b_{1|2}}\right)^{a_{12}}
\left(\frac{b_{1|2}(1+\tau^2_{12}) +(y_1-\mu_{1|2})^2}
{b_{2|1}(1+\tau^2_{12})+(y_2-\mu_{2|1})^2}\right)^{a_{12}+1}
\leq 1\label{numofnum}\\
{}&\Leftrightarrow 
\left(\frac{b_{2|1}}{b_{1|2}}\right)^{a_{12}}
\leq 1\nonumber\\
{}&\Leftrightarrow b_{2|1}-b_{1|2}
\leq 0{}&\nonumber\\
{}&:= \left(\beta +y_1^2+\mu^2/\tau^2-(\mu/\tau^2+y_1)^2\tau^2_{12}\right)\nonumber \\
{}&\hspace{1cm}-\left(\beta +y_2^2+\mu^2/\tau^2-(\mu/\tau^2+y_2)^2\tau^2_{12}\right) \leq0\nonumber\\
{}&\Leftrightarrow\left(y_1 -\mu/\tau^2 \frac{\tau^2_{12}}{1-\tau^2_{12}}\right)^2-\left(y_2 -\mu/\tau^2 \frac{\tau^2_{12}}{1-\tau^2_{12}}\right)^2\leq0\label{wanttomatch}
\end{align}

Since the second term in Equation \ref{numofnum} simplifies to 1:

\begin{align}
{}&\text{numerator}\left(\frac{b_{1|2}(1+\tau^2_{12}) +(y_1-\mu_{1|2})^2}
{b_{2|1}(1+\tau^2_{12})+(y_2-\mu_{2|1})^2}\right) \nonumber\\
:={}&b_{1|2}(1+\tau^2_{12}) +(y_1-\mu_{1|2})^2 \nonumber \\ \nonumber
:={}&\left(\beta  + {y_2}^2+\mu^2/\tau^2 -(\mu/\tau^2+{y_2})^2\tau_{12}^2 \right)(1+\tau_{12}^2) \\ \nonumber
{}&\hspace{1cm}+ ({y_1}-\left(\mu/\tau^2 +{y_2}\right)\tau_{12}^2)^2\\ \nonumber
{}&\hspace{1cm} +y_2 \left[ -2(\mu/\tau^2)\tau_{12}^2(1+\tau_{12}^2) +2(\mu/\tau^2)(\tau_{12}^2)^2 \right]\\ \nonumber
{}&\hspace{1cm} +y_1^2\\ \nonumber
{}&\hspace{1cm} + y_1 \left[-2(\mu/\tau^2)\tau_{12}^2\right]\\ \nonumber
{}&\hspace{1cm} + y_1y_2\left[-2\tau_{12}^2\right] \nonumber \\
={}&D + y_2^2 \label{Neqn}\\ 
{}&\hspace{1cm} +y_2 \left[ -2(\mu/\tau^2)\tau_{12}^2)  \right] \nonumber\\
{}&\hspace{1cm} +y_1^2 \nonumber\\ \nonumber
{}&\hspace{1cm} + y_1 \left[ -2(\mu/\tau^2)\tau_{12}^2\right]\\ \nonumber
{}&\hspace{1cm} + y_1y_2\left[ -2\tau_{12}^2\right],
\end{align}
for some function $D$ that does not depend on $y_1$ or $y_2$. 
Since the coefficients on the $y_1,y_{2}$ terms are the same, by symmetry, the denominator of the second term in Eqn \ref{numofnum} will also be equal to Equation \ref{Neqn}.

We now show the inequality  $C_B(\{y_1,y_2\},y_2)\leq C_B(\{y_1,y_2\},y_1)$ simplifies to Equation \ref{wanttomatch}. Notice that all parameters of the $t$ distribution that define this chosen conformity measure will be the same regardless of which variable we are obtaining the conformity score for. The conformity score is:
\[
C_B\left( \{y_1,y_2\}, X\right) = p(X | y_1,y_2)= dt_{\nu_{12}}\left(X|\mu_{12},{\variance_{12}}\right)
\]
where $\mu_{12} =  \left(\mu/\tau^2 +y_1 +y_2\right)\tau^2_{12'}$ and $\tau^2_{12'}=\left(1/\tau^2+2\right)^{-1}$ for any $X\in\mathcal{Y}$.

Then,
\begin{align}
{}&\hspace{7mm}C_B\left(\{y_1,y_2\},y_2\right) \leq C_B \left(\{y_1,y_2\},y_1\right) \nonumber\\
{}&\Leftrightarrow \left(y_1 - \left(\mu/\tau^2 +y_1 +y_2\right)\tau^2_{12'}\right)^2\leq \left(y_2- \left(\mu/\tau^2 +y_1 +y_2\right)\tau^2_{12'}\right)^2\nonumber\\
{}&\Leftrightarrow \left(y_1 - \mu/\tau^2 \frac{\tau^2_{12'}}{1-2\tau^2_{12'}}\right)^2- 
\left(y_2 - \mu/\tau^2 \frac{\tau^2_{12'}}{1-2\tau^2_{12'}}\right)^2\leq0\nonumber\\
{}&\Leftrightarrow \left(y_1 - \mu/\tau^2  \frac{\tau^2_{12}}{1-\tau^2_{12}}\right)^2- 
\left(y_2 - \mu/\tau^2  \frac{\tau^2_{12}}{1-\tau^2_{12}}\right)^2\leq0\label{matched}
\end{align}
since
\begin{align*}
\frac{\tau^2_{12'}}{1-2\tau^2_{12'}}:={}&\frac{\left(1/\tau^2+2\right)^{-1}}{1-2\left(1/\tau^2+2\right)^{-1}} \\
= {}& \frac{1}{\left(1/\tau^2+2\right)-2\left(1/\tau^2+2\right)\left(1/\tau^2+2\right)^{-1}}\\
= {}& \frac{1}{1/\tau^2}\\
= {}& \frac{1}{1/\tau^2 + 1 - \left( \tau^2_{12}\right)^{-1}\tau^2_{12}}\\
=: {}& \frac{1}{\left( \tau^2_{12}\right)^{-1}-\left( \tau^2_{12}\right)^{-1}\tau^2_{12}}\\
= {}& \frac{\tau^2_{12}}{1 - \tau^2_{12}}.
\end{align*}

In summary, by Equations \ref{wanttomatch} and \ref{matched} we have that
\begin{align*}
{}&\hspace{7mm}C_B\left(y_1,y_2\right) \leq C_B \left(y_2,y_1\right) \\
{}&\Leftrightarrow \left(y_1 - \mu/\tau^2  \frac{\tau^2_{12}}{1-\tau^2_{12}}\right)^2- 
\left(y_2 - \mu/\tau^2  \frac{\tau^2_{12}}{1-\tau^2_{12}}\right)^2\leq0\\
{}&\Leftrightarrow C_B\left(\{y_1,y_2\},y_2\right) \leq C_B \left(\{y_1,y_2\},y_1\right).
\end{align*}
Then, by Lemma \ref{CDequiv}, 
$C_B\left(\{Y_1,...,Y_{n+1}\}\backslash Y_i,Y_i\right)$ and $C_B\left(\{Y_1,...,Y_{n+1}\},Y_i\right)$ are ECD.
\end{proof}

\begin{proof}[Proof of Lemma \ref{smallinterval}]
As shown in the proof of Theorem \ref{conformitytheorem}, 
\begin{align*}
{}&\hspace{7mm}C\left(y_1,y_2\right) \leq C \left(y_2,y_1\right) \\
{}&\Leftrightarrow C\left(\{y_1,y_2\},y_2\right) \leq C \left(\{y_1,y_2\},y_1\right) \\
{}&\Leftrightarrow \left(y_1 - \mu/\tau^2  \frac{\tau^2_{12}}{1-\tau^2_{12}}\right)^2- 
\left(y_2 - \mu/\tau^2  \frac{\tau^2_{12}}{1-\tau^2_{12}}\right)^2\leq0
\end{align*}
Therefore, the inequality reduces to a quadratic function of the unknown $y_{1}$. Label this function $h$:
\[
h(y_1):=\left(y_1 - \mu/\tau^2  \frac{\tau^2_{12}}{1-\tau^2_{12}}\right)^2- 
\left(y_2 - \mu/\tau^2  \frac{\tau^2_{12}}{1-\tau^2_{12}}\right)^2\leq 0
\]
By standard quadratic formula theory, we can draw a few conclusions:
\begin{enumerate}
\item Notice that $h(y_1)$ is in vertex form. In $h(y_1)$, the coefficient on the first term is positive, so the parabola will be upward facing.  
\item The discriminant obtained via the quadratic formula is $4\left(y_2 - \mu/\tau^2  \frac{\tau^2_{12}}{1-\tau^2_{12}}\right)^2$. 
As the discriminant is non-negative, the solutions to $h(y_1)=0$
are a repeated real value or 2 unique real values.
\end{enumerate}
By items (1) and (2), the solution to the inequality $h(y_1)\leq 0$ will be an interval. 
Solving the inequality yields that the interval is of the form
\[
\left[\min\{y_2,g(y_2)\},\max\{y_2,g(y_2)\} \right]
\]
where $g(y_2) := (\mu/\tau^2)\tau^2_{12}(1-\tau^2_{12})^{-1}-y_2$. 
\end{proof}

\begin{proof}[Proof of Lemma \ref{containsshrinkage}]
Following notation used thus far in proofs, we aim to show
\[
\tilde{\theta}\in\left[\min\{y_2,g(y_2)\},\max\{y_2,g(y_2)\} \right]
\]
where $\tilde{\theta} := (\mu/\tau^2+y_2)\tau^2_{12}$ and $g(y_2) := (\mu/\tau^2)\tau^2_{12}(1-\tau^2_{12})^{-1}-y_2$. 
We first consider the case where $y_2<g(y_2)$. 

By the quadratic formula, 
we conclude the vertex $v$ of the parobola $h(y_1)$ is
\[
v:=(\mu/\tau^2)\tau^2_{12}(1-\tau^2_{12})^{-1}.
\]
and $y_2<v<f(y_2)$. 
We now prove the result in two steps.
\begin{enumerate}
\item First, we show the ordering $y_2<\tilde{\theta}$: 
\begin{align*}
{}&\hspace{7mm}y_2<(\mu/\tau^2)\tau^2_{12}(1-\tau^2_{12})^{-1} \\
{}&\Leftrightarrow y_2(1-\tau^2_{12})<(\mu/\tau^2)\tau^2_{12}\\
{}&\Leftrightarrow y_2<(\mu/\tau^2)\tau^2_{12}+y_2\tau^2_{12}\\
{}&=: y_2<\tilde{\mu}_\theta.
\end{align*}
\item Additionally, we can show $f(y_2)>\tilde{\theta}$:
\begin{align*}
{}&\hspace{7mm}y_2<\tilde{\theta}\\
{}&:=y_2<(\mu/\tau^2)\tau^2_{12}+y_2\tau^2_{12}\\
{}&\Leftrightarrow y_2\left(1-\tau^2_{12}\right)<(\mu/\tau^2)\tau^2_{12}\\
{}&\Leftrightarrow y_2<(\mu/\tau^2)\frac{\tau^2_{12}}{\left(1-\tau^2_{12}\right)}\\
{}&\Leftrightarrow y_2<(\mu/\tau^2)\left((1-\tau^2_{12})^{-1}-1\right)\\
{}&\Leftrightarrow y_2+\mu/\tau^2<(\mu/\tau^2)(1-\tau^2_{12})^{-1}\\
{}&\Leftrightarrow \left(y_2+\mu/\tau^2\right)\tau^2_{12}<(\mu/\tau^2)(1-\tau^2_{12})^{-1}\tau^2_{12}\\
{}&=: \tilde{\theta}<v
\end{align*}
since 
\[
(1-\tau^2_{12})^{-1}-1 = \frac{1-\left(1-\tau^2_{12}\right)}{1-\tau^2_{12}}= \frac{\tau^2_{12}}{1-\tau^2_{12}}
\]

To summarize, given that $y_2<f(y_2)$, we have shown:
\[
y_2<\tilde{\mu}_\theta<v_b<f(y_2)
\]
If $y_2>f(y_2)$, the ordering of the terms is reversed.
\end{enumerate}

\end{proof}

\begin{proof}[Proof of Thm \ref{maintheorem}]
By Lemmas \ref{smallinterval} and \ref{containsshrinkage}, each sub-region of acceptance $S_i$, for $i\in\{1,...,n+1\}$, is an interval which contains $\tilde{\theta}$.
Therefore, the hypothesis of Lemma \ref{stepwise} is met, and so, by Lemma \ref{interval}, the conformal prediction region is an interval which contains $\tilde{\theta}$.
Furthermore, by Equation \ref{boilsdown}, the conformal prediction region is
\begin{align*}
A^{fab}\left(\Ybf\right)
={}& \left\{y_{n+1}\in\mathcal{Y}:
\#\left\{i=1,...,n+1:c_i(y_{n+1}) \leq c_{n+1}(y_{n+1}) \right\}>k\right\},\\
\end{align*}
which is the $k$th and $(2n-k+1)$th order statistics of  the collection of bounds of the sub-regions of acceptance, $ \boldsymbol v = \begin{bmatrix}y_1&\cdots&y_n&g(y_1)&\cdots&g(y_n)\end{bmatrix}^T$
\end{proof}

\end{document}